\documentclass[manuscript]{acmart}
\AtBeginDocument{%
  }

\setcopyright{acmlicensed}
\copyrightyear{2018}
\acmYear{2018}
\acmDOI{XXXXXXX.XXXXXXX}

\graphicspath{{figs/}{figures/}{pictures/}{images/}{./}} 
\usepackage{cleveref}
\usepackage{tabu}                      
\usepackage{booktabs}                  
\usepackage{lipsum}                    
\usepackage{mwe}                       
\usepackage{mathptmx}
\usepackage{color}
\usepackage{fancyvrb}
\usepackage{xcolor}
\usepackage{array}
\usepackage{enumitem}
\usepackage{wrapfig}
\usepackage{longtable}

\usepackage{xspace,xpunctuate}

\newcommand{\etal}{\xspace\textit{et al.}\xspace}
\newcommand{\eg}{\textit{e.g.},\xspace}

\begin{document}

\title{Carbon and Silicon, Coexist or Compete? A Survey on Human-AI Interactions in Agent-based Modeling and Simulation}

\author{Ziyue Lin}
\email{ziyuelin917@gmail.com}
\orcid{0009-0002-5485-7379}
\affiliation{%
  \institution{School of Data Science, Fudan University}
   \streetaddress{1 Th{\o}rv{\"a}ld Circle}
  \city{Shanghai}
  \country{China}
}

\author{Siqi Shen}
\email{shensiqi.ssq@alibaba-inc.com}
\affiliation{%
  \institution{DataV Lab, Alibaba Group}
  \city{Hangzhou}
  \country{China}
}
\author{Zichen Cheng}
\email{zccheng19@fudan.edu.cn}
\affiliation{%
  \institution{School of Data Science, Fudan University}
  \city{Shanghai}
  \country{China}
}
\author{Cheok Lam Lai}
\email{thomaslai314159@gmail.com}
\affiliation{%
  \institution{School of Data Science, Fudan University}
  \city{Shanghai}
  \country{China}
}
\author{Siming Chen}
\authornote{Siming Chen is the corresponding author.}
\email{simingchen@fudan.edu.cn}
\orcid{0000-0002-2690-3588}
\affiliation{%
  \institution{School of Data Science, Fudan University}
  \city{Shanghai}
  \country{China}
}
\affiliation{%
  \institution{Shanghai Key Laboratory of Data Science}
  \city{Shanghai}
  \country{China}
}
\renewcommand{\shortauthors}{Lin et al.}

\begin{abstract}
Recent interest in human-AI interactions in agent-based modeling and simulation~(ABMS) has grown rapidly due to the widespread utilization of large language models~(LLMs).
ABMS is an intelligent approach that simulates autonomous agents' behaviors within a defined environment to research emergent phenomena.
Integrating LLMs into ABMS enables natural language interaction between humans and models. 
Meanwhile, it introduces new challenges that rely on human interaction to address.
Human involvement can assist ABMS in adapting to flexible and complex research demands.
However, systematic reviews of interactions that examine how humans and AI interact in ABMS are lacking.
In this paper, we investigate existing works and propose a novel taxonomy to categorize the interactions derived from them.
Specifically, human users refer to researchers who utilize ABMS tools to conduct their studies in our survey. 
We decompose interactions into five dimensions: the goals that users want to achieve~(Why), the phases that users are involved~(When), the components of the system~(What), the roles of users~(Who), and the means of interactions~(How).
Our analysis summarizes the findings that reveal existing interaction patterns.
They provide researchers who develop interactions with comprehensive guidance on how humans and AI interact.
We further discuss the unexplored interactions and suggest future research directions.
\end{abstract}

\ccsdesc[500]{Human-centered computing~Interactive systems and tools}

\keywords{agent-based modeling and simulation, human-AI interactions}

\begin{teaserfigure}
  \includegraphics[width=\textwidth]{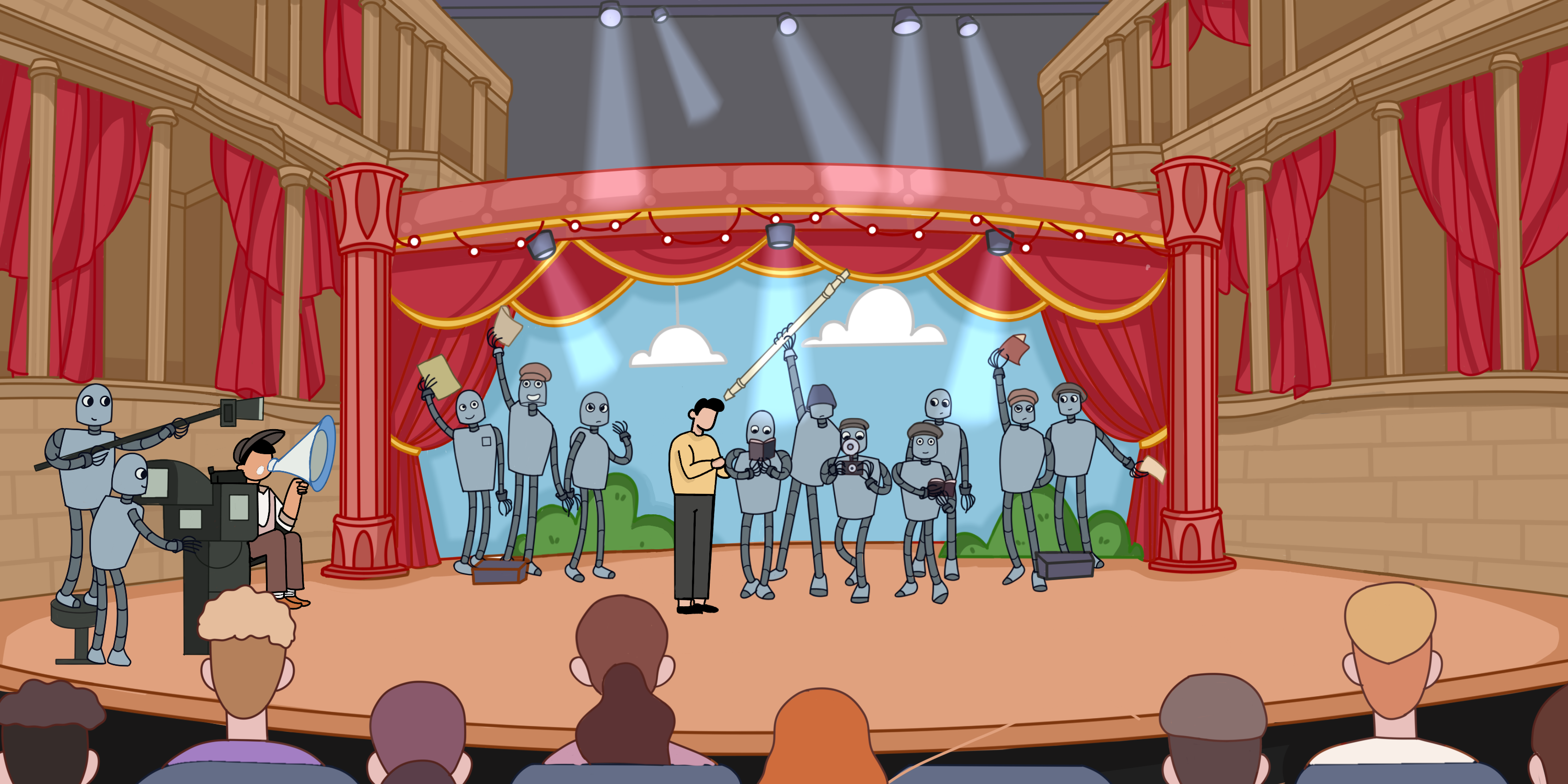}
  \caption{Scenario of an envisioned agent-based modeling and simulation composed of agents, in which humans can also participate. Human-AI interactive ABMS can be effectively explained through an analogy from the field of theater. The above image depicts agents as actors on stage, while humans can take on roles such as director, actor, observer, etc.}
  \label{fig:teaser}
\end{teaserfigure}

\maketitle
\section{Introduction}

Agent-based modeling and simulation~(ABMS)~\cite{1574234} has long been recognized as a powerful approach for studying complex systems~\cite{gilbert2004agent} in various domains, including sociology~\cite{annurev:/content/journals/10.1146/annurev.soc.28.110601.141117,gilbert_how_2000}, economics~\cite{hamill2015agent, LENGNICK2013102}, ecology~\cite{MCLANE20111544}, and epidemiology~\cite{el-sayed_social_2012}.
ABMS is a computational approach to model complex systems composed of autonomous agents.
By allowing researchers to simulate the behaviors of individual agents within an extensive system, ABMS enables the exploration of emergent phenomena that arise from these behaviors~\cite{AN201225}.
This capability to model complex, dynamic systems has made ABMS indispensable in understanding collective behaviors and testing scenarios in environments that would be challenging or impossible to replicate in reality~\cite{heath2009}.
As artificial intelligence~(AI) continues to advance, the integration of human-AI interaction within ABMS offers significant potential to enhance ABMS's applicability~\cite{doi:10.1177/0037549706073695,berryman2008review}.
Human users interact with models to customize them based on their specific research requirements~\cite{netlogo}, such as adjusting parameters, guiding agent behaviors, or testing new scenarios to adapt models on the fly~\cite{10.1145/3613904.3642545,10.1145/3490099.3511105}.
This level of interaction opens doors to more accurate, adaptive, and user-centered simulations.

The emergence of large language models~(LLMs)~\cite{bommasani2022opportunities,brown_language_2020} has further expanded the potential of ABMS by facilitating more natural and intuitive human-AI interactions. 
Gao\etal~\cite{10.1145/3613905.3650786} have illustrated that human-LLM interaction facilitates more complex reasoning and creativity tasks.
With LLMs, users can communicate with the simulation through natural language, lowering the barrier to entry for non-experts without programming skills and enhancing the user experience.
Park\etal~\cite{10.1145/3586183.3606763} proposed Generative Agent that supports users creating agents and communicating with agents directly by natural language.
The user-friendly design has significantly advanced the interactivity of ABMS, spurring new applications and fostering interdisciplinary research~\cite{DBLP:journals/corr/abs-2312-11813,10.1145/3613904.3642159}.
Meanwhile, it presents new challenges for ABMS, necessitating solutions through human-AI interactions. 
One key challenge lies in evaluating the effectiveness of these outcomes, as traditional statistical metrics often fall short of capturing the complexity and nuances of agent behaviors~\cite{10.1145/3526113.3545616, doi:10.1126/science.ade9097}. 

The combination of enhanced human-AI interactivity and the accessibility brought by LLMs has attracted a growing number of researchers from diverse fields, including HCI, AI, ubiquitous computing, and social science, to explore the potential of ABMS. 
This renewed interest and broadened expertise contribute to a rapidly evolving landscape, pushing the boundaries of ABMS beyond traditional applications and expanding its relevance to novel, interdisciplinary challenges.
Developers of ABMS are beginning to explore how the design of interactions can enhance ABMS to serve user research needs better.
However, designing effective human-AI interactions is not trivial.
On the one hand, the inherent complexity of ABMS itself requires interaction methods that can adapt to dynamic, often non-linear, changes within the simulation.
On the other hand, enabling effective communication between users and models is challenging, as it requires real-time feedback mechanisms that facilitate clear understanding and support decision-making.
Existing surveys on ABMS leveraging LLMs~(\eg~\cite{gao_large_2023}) have not focused on summarizing human-AI interactions.
There still is a lack of systematic surveys on human-AI interactions in ABMS to provide an overview of the research landscape.
This paper seeks to address the following research question to support reflection on current research progress and future opportunities: \textit{How do humans and AI interact in the context of ABMS to fulfill user research requirements?} 
We address this question from two perspectives: what research goals users aim to achieve through interaction and how to design specific interactions once the goals are determined. 

To fill this research gap, we first collected 97 relevant studies about human-AI interactions in ABMS.
We extracted human-AI interactions from the papers in the corpus.
Our survey defines human users as researchers employing ABMS tools to conduct their studies.
We decomposed each interaction into five dimensions according to our taxonomy framework, which is derived from the "5W1H" guideline~\cite{ram_5ws_2018}.
The five dimensions are: \textit{Why}, \textit{When}, \textit{What}, \textit{Who}, \textit{How}.
\textit{Why} explains the motivations of users.
Users find it challenging to accomplish their goals with static models, making interactions essential.
\textit{When} refers to the phase at which users are involved in the simulation.
\textit{What} pertains to the components of the system under user control. Considering the simulation system's characteristics, \textit{What} encompasses three primary aspects of the model: agents, environment, and simulation configuration.
\textit{Who} represents the roles that users assume during the interaction process.
We draw an analogy from theater, where the behavior of agents within the model mirrors the actions of actors performing on stage.
\textit{How} refers to the means employed by users to interact with the model.
By integrating five dimensions, we can comprehensively understand the design of human-AI interactions in ABMS.

The papers we examined span a wide timeframe, from 1996 to 2024, and cover multiple fields, including human-computer interaction, ubiquitous computing, natural language processing, computer vision, political science, sociology, and more.
Human-AI interactions range from scientific simulation software platforms to more flexible and diverse modes of user engagement.
Significantly, as LLMs lower the barrier to interaction, they have attracted many AI and HCI researchers to engage in related studies. 
This development expands the application scope and potential of ABMS, extending beyond merely simulating macro-level collective behaviors or phenomena.
We summarized comprehensive findings illuminating existing interaction patterns within ABMS, revealing established trends and frameworks, and identifying critical gaps in current research.
We hope our study can suggest directions for future research that can guide the developers of ABMS in developing more effective, user-centered, and versatile interactive systems.
In summary, our main contributions to the domain are as follows:
\begin{itemize}
\item We present the first comprehensive survey on human-AI interactions in agent-based modeling and simulation and introduce a novel taxonomy of interactions derived from an extensive review of existing literature.
\item We synthesize the findings from existing literature using our proposed taxonomy, which reveals interaction patterns, highlights research gaps, and suggests future research directions.
\end{itemize}
\section{Background}
This section discusses relevant studies about agent-based modeling and simulation~(ABMS) and human-AI interaction for ABMS.
\subsection{Agent-based Modeling and Simulation}
Autonomous agents demonstrate varying degrees of intelligence, enabling them to perceive their environment, make decisions, and execute actions in pursuit of certain goals~\cite{10.1007/BFb0013570,wooldridge_jennings_1995}.
Agent-based modeling and simulation~(ABMS) connects the micro-level actions of individual agents to the macro-level dynamics of the overall system~\cite{8352646}.
The investigation of ABMS has been a longstanding area of focus within the field of artificial intelligence research~\cite{1574234,5429318,doi:10.1177/0037549706073695}.
ABMS is a powerful method for simulating complex social systems. 
It constructs a computational environment to allow autonomous, dynamic, and heterogeneous agents to interact with one another and their surroundings, acting according to predefined rules or behaviors.
This approach enables the exploration of emergent phenomena arising from individual agent interactions within the complex system~\cite{helbing_agent-based_2012,doi:10.1073/pnas.072081299}.
ABMS demonstrates remarkable flexibility and has been utilized in a wide range of disciplines.
The global financial system ranks as one of the most intricate systems developed by humans~\cite{Wang_2018, Samanidou_2007}.
Ponta\etal~\cite{Ponta_2011} presented a multi-asset, agent-based financial market model composed of zero-intelligence agents with limited financial resources.
Random allocation strategies were employed for agents constrained by their finite resources. 
The resulting stock market dynamics exhibit stylized facts, including volatility clustering, fat-tailed return distributions, and mean reversion tendencies.
ABMS can help public healthcare administrators identify interventions that enhance population wellness and quality of care while concurrently reducing costs~\cite{SILVERMAN201561,williams2023epidemicmodelinggenerativeagents,el-sayed_social_2012}.
Researchers and public health officials across many countries have utilized Covasim~\cite {kerr_covasim_2021} to forecast epidemic trends, evaluate intervention scenarios, and assess resource requirements~\cite{SILVA2020110088}.
Furthermore, Conte\etal~\cite{10.3389/fpsyg.2014.00668} presented that interdisciplinary computational social science uses ABSM to verify internal consistency, examine the resulting aggregate states, and employ cross-methodological experimental approaches to validate hypotheses against real-world data.
With the advancement of Internet technology, social media has transformed our way of life~\cite{KAPLAN201059}.
Gatti\etal~\cite{gatti_large-scale_2014} proposed stochastic multi-agent-based modeling to simulate what users post on an egocentric social network, where Barack Obama is considered as the central user.

The emergence of powerful capabilities in large language models~(LLMs)~\cite{bommasani2022opportunities,brown_language_2020} enables agents to exhibit more human-like behaviors~\cite{NBERw31122}, sparking significant interest in ABMS among an increasing number of researchers from AI and HCI community.
In contrast to predefined rules and decision trees~\cite{marcotte_behavior_2017}, LLMs add flexibility by allowing agents to adapt and respond to complex situations dynamically, improving the quality of behavioral modeling.
Park\etal~\cite{10.1145/3586183.3606763}, leveraging LLMs' power, introduced generative agents that can simulate believable human behaviors with architecture for synthesizing and retrieving relevant information.
Moreover, LLMs broaden the application scenarios for ABMS by enabling more nuanced, human-like agent interactions and expanding the scope of dynamic environments that can be realistically modeled.
AGENTVERSE~\cite{chen2023agentversefacilitatingmultiagentcollaboration} emphasized the effectiveness of the multi-agent collaboration on text understanding, coding, and tool utilization.
$S^3$~\cite{gao2023s3socialnetworksimulationlarge} simulated social networks with LLM-empowered agents to capture three forms of propagation: information, emotion, and attitude.
Chatlaw~\cite{cui2024chatlawmultiagentcollaborativelegal} applied a multi-agent system to improve the reliability and precision of AI-powered legal services.
A chat-powered software development framework where LLMs power specialized agents to design, code, and test software~\cite {qian2024chatdevcommunicativeagentssoftware}.
There exist previous surveys on LLM-empowered agents~\cite{xi2023risepotentiallargelanguage,wang_survey_2024} and ABMS~\cite{gao_large_2023}.
Their primary focus is on how to design simulation agents and how to build simulation environments.
Although Xi\etal~\cite{xi2023risepotentiallargelanguage} summarized two paradigms of human-agent interaction, there is a lack of systematic surveys to investigate how humans interact with the ABMS system.
To fill the gap, we first categorized human-AI interactive methods in the context of ABMS according to the ``5W1H'' guideline.

\subsection{Human-AI Interaction for ABMS}
The development of the ABMS scientific simulation platform has evolved over several decades~\cite{doi:10.1177/0037549706073695,berryman2008review}, driven by advances in computational power, the need for more realistic modeling of complex systems, and interdisciplinary applications for researchers. 
These platforms provide the frameworks and tools that allow users to build, run, and analyze ABMS across various domains.
For example, MASON~\cite{doi:10.1177/0037549705058073} is a high-performance agent-based simulation toolkit developed in Java, allowing users to build complex models by combining customizable components and handle simulations involving thousands of agents efficiently.
NetLogo~\cite{netlogo} is another high-level platform offering a simple yet robust programming language, integrated graphical interfaces, and extensive documentation.
It is mainly designed for ABMS involving dynamic individuals with local interactions within a grid space.
Although NetLogo’s custom language is user-friendly, it is limited in functionality and flexibility compared to Python or Java, which restricts the depth of customization available to advanced users or those needing complex system handling and processing capabilities.
Guyot\etal~\cite{guyot2006} proposed ``agent-based participatory simulations'' methods to simulate multi-agent systems where human participants can control some of the agents.
Furthermore, real human demographic information can be utilized for the initialization of ABSM systems~\cite{GAUBE201392,10.1145/3394486.3412862}.
Human-AI interactions in ABMS have also been studied in Role-Playing Games~(RPGs), a genre of games where players assume the roles of characters in a fictional world, interacting within a narrative-rich environment~\cite{riedl_interactive_2021}.
Autonomous agents are well known for appearing in these games as non-player characters~(NPCs)~\cite{10.1145/2282338.2282384,10.1145/2159365.2159425}.
These games and agents are designed to offer immersive experiences, allowing players to engage in character development, story progression, and tactical or strategic gameplay~\cite{brenner_creating_2010,isbister_consistency_2000}.

The natural language capabilities of LLMs lower the technical barriers for ABMS users, allowing those without extensive programming skills to design and adjust simulations through conversational commands.
LLMs offer expanded possibilities for interactions, broadening the boundaries and applications of ABMS.
Park\etal~\cite{10.1145/3586183.3606763} presented the system that supports users defining, controlling, and intervening agents and environments by natural language commands. 
Memory Sandbox~\cite{10.1145/3586182.3615796} allows users to manage agents' memories to align with users' understandings by interface and conversations.
Leveraging the vast datasets used to train LLMs, agents can display diverse behaviors that reflect distinct characteristics, enhancing the realism and variety of simulated results.
It has introduced new challenges for ABMS, which require resolution through human-AI interactions.
For example, it is difficult to evaluate the effectiveness of the outcomes using traditional statistical metrics.
Researchers have sought to assess ABMS through methodologies grounded in human-AI interaction.
Social Simulacra~\cite{10.1145/3526113.3545616} recruited human participants to evaluate the believability of simulated behaviors by asking whether they could distinguish a conversation generated by either humans or agents.
To evaluate the ability to play a strategy game involving both cooperation and competition, Cicero~\cite{doi:10.1126/science.ade9097} participated anonymously in 40 games with humans on the website and placed first in this tournament.
To study the trends of interactive patterns between humans and AI, we collected relevant literature and introduced a novel taxonomy on interactions.
Existing work summarized the taxonomy of LLM-human interaction modes~\cite{10.1145/3613905.3650786}.
Inspired by the categorization, we decompose the human-AI interactive methods of ABMS into five dimensions: \textit{Why}, \textit{When}, \textit{What}, \textit{Who}, and \textit{How}.
\section{Methodology}
This section introduces our method of collecting and coding the corpus of works.
Then, we provide the findings of the descriptive statistics concerning publication year, publication venue, and prominent works.
\subsection{Paper Collection}
We applied two kinds of methods, reference-driven and search-driven, to collect relevant papers and research.
First, we collected papers within the scope of our survey from the latest core literature reviews about related topics: ABMS~\cite{gao_large_2023}, LLM-empowered agents~\cite{xi2023risepotentiallargelanguage,wang_survey_2024}, and human-LLM interactions~\cite{10.1145/3613905.3650786}.
Second, we developed our search query based on the collected papers.
We summarized the keyword list: ``agent'',  ``large language models'', ``GPT'', ``LLMs'', ``interaction'' and ``human-AI''.
Since the literature reviews we refer to were published in 2023 or 2024, we focus on papers in these two years during the search phase.
We further expanded the corpus by including papers that either cited these works or were referenced by them within the corpus.
Subsequently, three co-authors separately reviewed all the works and filtered those that fall within the scope of our research.
We established two filtering criteria for the corpus due to the implicit search keywords.
First, the paper related to LLMs should involve research on agents that simulate human behaviors, rather than merely exploring the capabilities of LLMs.
Second, the papers must address human-AI interactions in ABMS, not just a static ABMS model.
Specifically, when humans are mentioned, they should refer to users of ABMS rather than interaction developers.
Disagreements regarding paper selection were addressed through multiple rounds of discussions among the three co-authors.
Eventually, we collected 97 works for further analysis in this paper.

\subsection{Descriptive Statistics}
We display descriptive statistics with respect to publication year, venue, and prominent work, which provides an overview of our corpus of works.

\begin{figure}[ht]
  \centering
  \includegraphics[width=\linewidth]{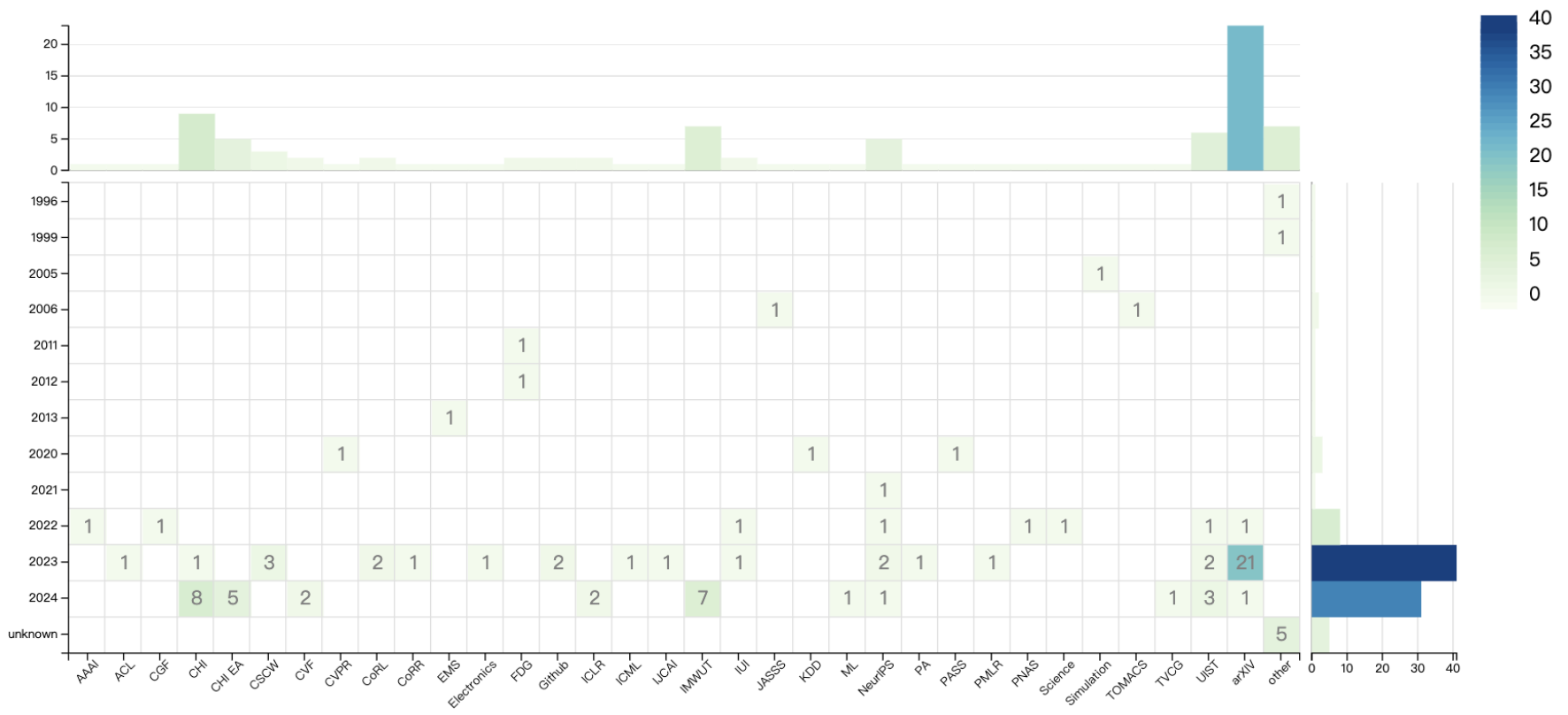}
  \caption{%
    \textbf{The statistical figure of publication year and venue.} Some venue names are abbreviated: Environmental Modelling \& Software~(EMS), Mind \& Language~(ML), Political Analysis~(PA), Proceedings of the Annual Simulation Symposium~(PASS). }
  \label{fig:stat}
\end{figure}

\subsubsection{Publication Year}
We first conducted a statistical analysis of the publication year of relevant papers. The first notable work in the field was published in 1996 by Minar\etal~\cite{minar1996swarm}, introducing a multi-agent software platform designed for the simulation of complex adaptive systems. Publication frequency remained relatively low until 2021. With the emergence and widespread adoption of LLMs, ABMS was empowered to facilitate more natural and intuitive interactions. A marked increase in the number of published papers was observed in 2022, followed by a dramatic surge in 2023, where 73\% of the articles were published thereafter. Research interest remains high in 2024, as shown in Fig~\ref{fig:stat}.
Several works lacking year information are early-developed simulation software or toolkits, with publication years unavailable.

\subsubsection{Publication Venue}
In terms of publication venues, we categorize 32 different venues into three major groups: the AI community~(\eg NeurIPS, EMNLP, AAAI), the HCI community~(\eg CHI, CSCW, UIST, IMWUT), and Others. To ensure the timeliness, quite a large amount of papers~(25.8\%) are collected from arXiv, reflecting the latest advancements and trends in the field. While the ACM CHI Conference on Human Factors in Computing Systems~(CHI)~(15.7\%) is the most common venue to appear for the journal/conference publications (including those in CHI EA), followed by IMWUT~(7.2\%), the Annual ACM Symposium on User Interface Software and Technology~(UIST)~(6.7\%) and the Conference on Neural Information Processing Systems~(NeurIPS)~(5.6\%).

\subsubsection{Prominent Work}

\begin{table}[ht]
  \caption{Most cited papers of interactive ABMS~(Top 10)}
  \label{tab:freq}
  \begin{tabular}{>{\arraybackslash}p{7.5cm} p{3cm} l l}
    \toprule
    \textbf{Title and Year} & \textbf{Authors} & \textbf{Venue} & \textbf{Citations} \\
    \midrule
    \specialrule{0em}{2pt}{2pt}
    Generative Agents: Interactive Simulacra of Human Behavior (2023) & Park\etal~\cite{10.1145/3586183.3606763} & UIST & 1636\\
    \specialrule{0em}{2pt}{2pt}
    MASON: A Multiagent Simulation Environment (2005) & Luke\etal~\cite{doi:10.1177/0037549705058073} & Simulation & 1444\\
    \specialrule{0em}{2pt}{2pt}
    Do As I Can, Not As I Say: Grounding Language in Robotic Affordances (2023) & Ahn\etal~\cite{ahn2022icanisay} & arXiv & 1242\\
    \specialrule{0em}{2pt}{2pt}
    The Swarm Simulation System: A Toolkit for Building Multi-Agent Simulations (1996) & Minar\etal~\cite{minar1996swarm} &- & 1201\\
    \specialrule{0em}{2pt}{2pt}
    Experiences creating three implementations of the repast agent modeling toolkit (2006) & North\etal~\cite{10.1145/1122012.1122013} & TOMACS & 941\\
    \specialrule{0em}{2pt}{2pt}
    Inner Monologue: Embodied Reasoning through Planning with Language Models (2022) & Huang\etal~\cite{huang2022innermonologueembodiedreasoning} & arXiv & 745\\
    \specialrule{0em}{2pt}{2pt}
    ALFRED: A Benchmark for Interpreting Grounded Instructions for Everyday Tasks (2020) & Schridhar\etal~\cite{shridhar2020alfredbenchmarkinterpretinggrounded} & CVPR & 732\\
    \specialrule{0em}{2pt}{2pt}
    MetaGPT: Meta Programming for A Multi-Agent Collaborative Framework (2023) & Hong\etal~\cite{hong2024metagptmetaprogrammingmultiagent} & arXiv & 414\\
    \specialrule{0em}{2pt}{2pt}
    Out of One, Many: Using Language Models to Simulate Human Samples (2023) & Argyle\etal~\cite{Argyle_Busby_Fulda_Gubler_Rytting_Wingate_2023} & Political Analysis & 414\\
    \specialrule{0em}{2pt}{2pt}
    CAMEL: Communicative Agents for "Mind" Exploration of Large Language Model Society (2023) & Li\etal~\cite{NEURIPS2023_a3621ee9}& NeurIPS & 404\\
  \bottomrule
\end{tabular}
\end{table}

We also assessed the influence of the included papers by examining their citation counts~(\Cref{tab:freq}). Then, we ranked the papers according to their citation counts and found that a large portion of these articles appeared after 2020 ($n=7$ in top 10). The rapid rise of LLMs during that time might be a possible reason. 
The most influential work is the 2023 paper Generative Agents~\cite{10.1145/3586183.3606763}. It introduces Generative Agents to simulate realistic human behaviors for interactive applications. The simulation was validated in a virtual small-town setting, and the agents successfully exhibited realistic individual and emergent social behaviors.
Users are extensively engaged in various ways throughout the simulation process via natural language, significantly reducing the learning costs associated with interaction methods.
Another significant work, MASON~\cite{doi:10.1177/0037549705058073} in our corpus, was cited 1444 times by Jan 2025. It introduces a Java-based, discrete-event simulation toolkit, which aims to provide a flexible, fast, and extensible simulation environment that separates the simulation model from visualization. 
Other highly cited works are mostly about overcoming the weakness of language models and enabling models to accomplish more complex tasks~\cite{ahn2022icanisay, shridhar2020alfredbenchmarkinterpretinggrounded, hong2024metagptmetaprogrammingmultiagent}.
\subsection{Paper Coding}
This paper aims to categorize the interactions derived from existing works.
Inspired by the taxonomy of human-LLM interaction modes proposed by Gao\etal~\cite{10.1145/3613905.3650786}, we adapted ``5W1H'' guideline~\cite{ram_5ws_2018} to decompose interactive methods between human and AI.
Following an initial review of all the papers in our corpus, we established a preliminary framework for paper coding.
Two co-authors coded the corpus separately based on both papers and related demos or presentation videos.
Next, the co-authors checked conflict coding and articulated their perspectives.
They modified the coding and refined the framework iteratively until diverging opinions were resolved.
For each paper or work, we extracted interactive methods from it and analyzed them within our framework.
Taking Generative Agents~\cite{10.1145/3586183.3606763} as an example,
we identified seven types of interactive methods in this paper.
We decomposed each interactive method into five dimensions: why, when, what, who, and how, according to our framework.
We presented the details of our framework in Section~\ref{framework}.

\section{Framework}~\label{framework}
\begin{figure}[htp]
  \centering
  \includegraphics[width=\linewidth]{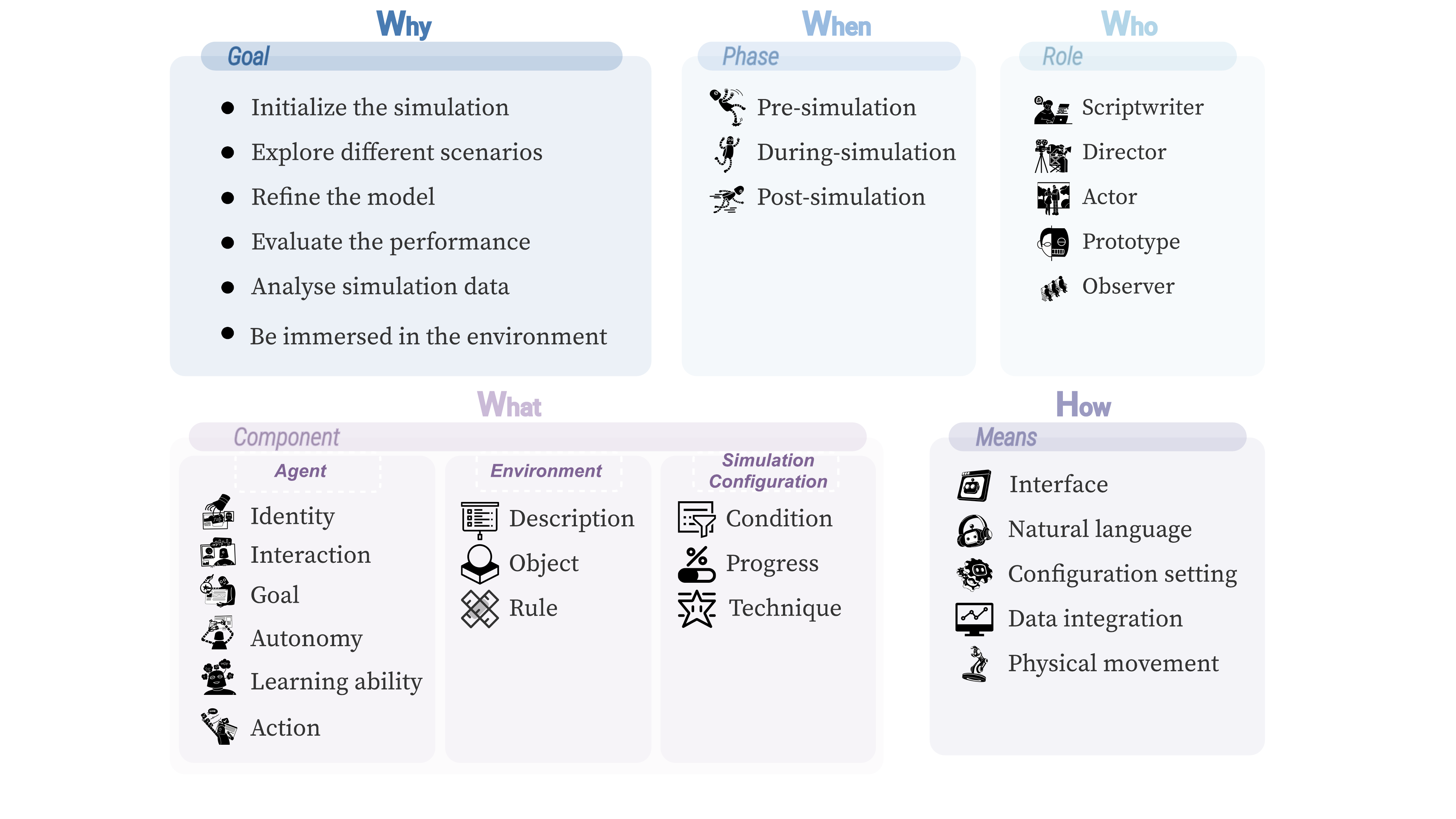}
  \caption{%
    \textbf{The details of our taxonomy.} We have five key dimensions to construct our taxonomy: the goals that users want to achieve~(Why), the phases that users are involved~(When), the roles of users~(Who), the components of the system~(What), and the means of interactions~(How).}
  \label{fig:tax}
\end{figure}

This section introduces the framework~(Fig~\ref{fig:tax}) for characterizing the interactions derived from the collected research.
Initially, we introduced the overview of our framework about applying the ``5W1H'' guideline~\cite{ram_5ws_2018} to decompose interactions.
Then, we provided detailed information about the dimensions of ``5W1H''.
Through these interactions, users can push the boundaries of ABMS, catering to personalized research needs.

\subsection{Types of Interactions}
Inspired by the taxonomy for human-LLM interactions proposed by Gao\etal~\cite{10.1145/3613905.3650786}, we adapted the ``5W1H'' guideline to categorize interactive methods from existing works.
Through an analysis of the characteristics of interactive methods, we have selected five key dimensions to construct our taxonomy:
\begin{itemize}
\item {\textbf{Why}}: the reasons or motivations behind the interactions. The goals users aim to achieve through interactive methods are difficult to accomplish with static models.
We summarize six goals: initialize the simulation, explore different scenarios, refine the model, evaluate the performance, analyze simulation data, and be immersed in the environment.
\item {\textbf{When}}: the phase at which users are involved in the simulation. 
We have divided it into three phases: pre-simulation~\cite{gao2023s3socialnetworksimulationlarge}, during-simulation~\cite{chen2023agentversefacilitatingmultiagentcollaboration,Padmakumar_Thomason_Shrivastava_Lange_Narayan-Chen_Gella_Piramuthu_Tur_Hakkani-Tur_2022}, and post-simulation~\cite{10520238}.
\item {\textbf{What}}: the components of the system controlled by users. 
Based on the features of the simulation system, we consider three main aspects of the model: agents, environment, and simulation configuration.
Additionally, we perform a secondary classification based on the three aspects, with the specific details explained in Section~\ref {what}.
\item {\textbf{Who}}: the roles users play during the interaction process. 
In this context, we employ an analogy from the field of theater, as the behavior of agents within the model parallels the actions of actors performing in a theatrical setting.
Therefore, we draw upon some related professions to correspond to the roles of users engaged in the model: scriptwriter, director, actor, prototype, and observer.
\item {\textbf{How}}: the means employed by users to interact with the model. 
We categorize it into interface, natural language, configuration setting, data integration, and physical movement.
\end{itemize}

Subsequently, we will provide a detailed description of the four dimensions, excluding the ``When'' dimension. The taxonomy framework is also shown in Figure~\ref{fig:tax}.

\subsection{Why: Classification of Goals}\label{goal}
After reviewing all the literature, we identified six goals that encapsulate the multifaceted role of human engagement in shaping ABMS.
They drive users to interact with the model since a non-interactive model may fail to align with the users' requirements sufficiently.

\textit{Initialize the Simulation.}
The first step, where users lay the foundation for the simulation, ensures that it aligns with the study's objectives.
Users can initiate the simulation by determining factors such as agents' characteristics, environmental variables, and simulation conditions~\cite{10.1145/3586183.3606763, 10.1145/3526113.3545616}.
Besides, users can decide when the simulations begin by issuing a start command~\cite{chen2023agentversefacilitatingmultiagentcollaboration,chan2023chatevalbetterllmbasedevaluators} and posing specific questions or requirements \cite{ren2023robotsaskhelpuncertainty}.
Typically, this step requires the users to incorporate domain-specific knowledge to set up the environment and populate the model with agents that reflect real-world entities or phenomena~\cite{GAUBE201392}.
With users' cooperation, the setup requirements for model initialization are met, laying the groundwork for the simulation to run.

\textit{Explore Different Scenarios.}
One of the critical goals for users in ABMS is to explore various hypothetical scenarios by adjusting key parameters.
It facilitates users exploring how different assumptions or interventions may impact system dynamics~\cite{10.1145/3613904.3642159}.
It also enables the discovery of insights that may not be immediately apparent from the initial model configuration.
Users can conduct "what-if" analysis and test multiple hypotheses in real-time by simulating alternative futures to uncover patterns that are otherwise difficult to detect in a static model~\cite{10.1145/3526113.3545616}.
The iterative process allows for a more thorough analysis of potential risks and opportunities in the modeled system~\cite{hua2024warpeacewaragentlarge}.


\textit{Refine the Model.}
If the model's performance falls short of expectations, user intervention is required to improve its effectiveness.
For example, agent behaviors may not align with observed real-world outcomes perfectly due to the simplification of action rules or limitations of the algorithmic capabilities.
To improve the relevance of the simulation results, users can make enhancements or corrections directly through interaction methods~\cite{mandi2023rocodialecticmultirobotcollaboration}.
Additionally, the learning abilities of agents can be improved through user involvement by providing learning materials or managing the agents' 
memory~\cite{jin2024surrealdriverdesigningllmpoweredgenerative,unknown}.
Users can also directly collaborate with agents in solving tasks or guide agents with instructions\cite{mohanty2023transforminghumancenteredaicollaboration,zhang2024buildingcooperativeembodiedagents}.
Due to the randomness inherent in some simulation algorithms, users can refine the model simply by regenerating the results~\cite{10.1145/3526113.3545616}.
The ability to refine models ensures the model's predictive power and validity, leading to more robust and sophisticated outcomes.

\textit{Evaluate the Performance.}
Humans play a central role in evaluating the performance of the ABMS by assessing how well the simulation meets predefined goals, such as accurately representing system dynamics, producing meaningful results, or predicting real-world behaviors.
By integrating user-centered metrics, this evaluation typically goes beyond standard quantitative measures~(\eg accuracy, speed).
Furthermore, qualitative feedback by users who incorporate domain-expert knowledge and subjective insights is essential, particularly in the era of LLMs.
Users can assess whether agent behaviors accurately simulate human actions by applying common sense or domain-specific knowledge~\cite{10.1145/3526113.3545616,10.1145/3613905.3651026,wang2023humanoidagentsplatformsimulating}.
It is significant for users to evaluate how effectively the model adapts to different contexts or scenarios and handles changes in user goals, external factors, or input variations~\cite{10.1145/3613905.3651008,10.1145/3613904.3642947}.

\textit{Analyze Simulation Data.}
Analyzing data generated by agent-based modeling and simulation is another critical goal for users.
Simulation data provides the foundation for understanding system behaviors, validating models, and making informed decisions.
Users can observe emergent patterns and system dynamics that may be difficult or impossible to study in the real world~\cite{berryman2008review}.
Furthermore, users employ statistical techniques~\cite{netlogo}, visual analytics~\cite{pan2024agentcoordvisuallyexploringcoordination,10520238}, and domain expertise~\cite{electronics12122722} to extract meaningful insights from the data, which can inform decision-making, strategy recommendations, or further model adjustments. 


\textit{Be Immersed in the Environment.}
In contrast to the goals mentioned above, being immersed in the environment emphasizes the user's experience within the simulation without the primary focus being on control or modification.
The focus is less on achieving a specific objective and more on how deeply the user engages with and experiences the simulation.
Direct interactions allow users to engage with agents' worlds actively, enhancing their entertainment experience~\cite{10.1145/3586183.3606763}.
It is most evident in mediums such as video games, virtual reality~(VR), and augmented reality~(AR), where users can fully immerse themselves in dynamic, interactive environments~\cite{mao2024alympicsllmagentsmeet}.


\subsection{What: Components of System}\label{what}
By analyzing the structure of the model, we detailed the components that users can control, focusing on three primary aspects: agents, environment, and simulation configuration.
\subsubsection{Agents} In ABMS, agents are often designed with human-like characteristics to simulate behaviors that closely mimic real-world scenarios. 
Agents are diverse, heterogeneous, and dynamic due to the complex components being divided into internal states and outward behaviors.
Based on the certain characteristics of agents proposed by Macal\etal~\cite{1574234}, we summarized five key components of internal states as follows:

\begin{itemize}
\item {\includegraphics[width=0.05\textwidth]{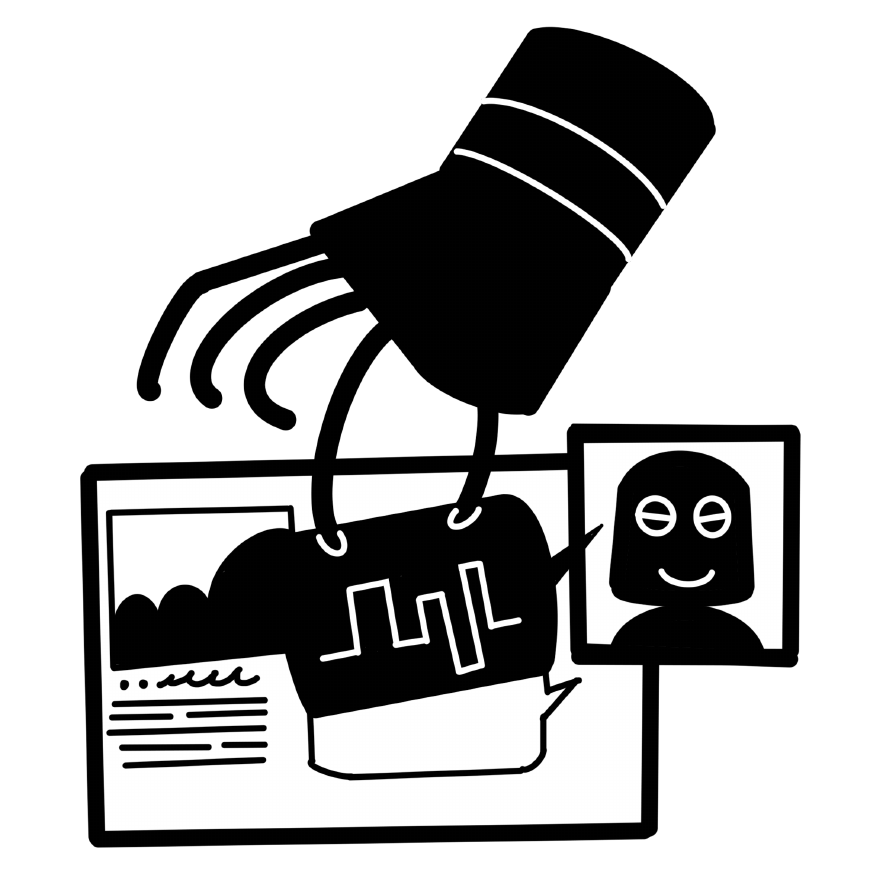}\textbf{Identity}}: We considered agents as discrete individuals with a set of attributes and rules~\cite{10.1145/3626772.3657844}. 
Agents can be endowed with human-like traits or specific behavioral abilities and rules~\cite{wang2024userbehaviorsimulationlarge}.

\item {\includegraphics[width=0.05\textwidth]{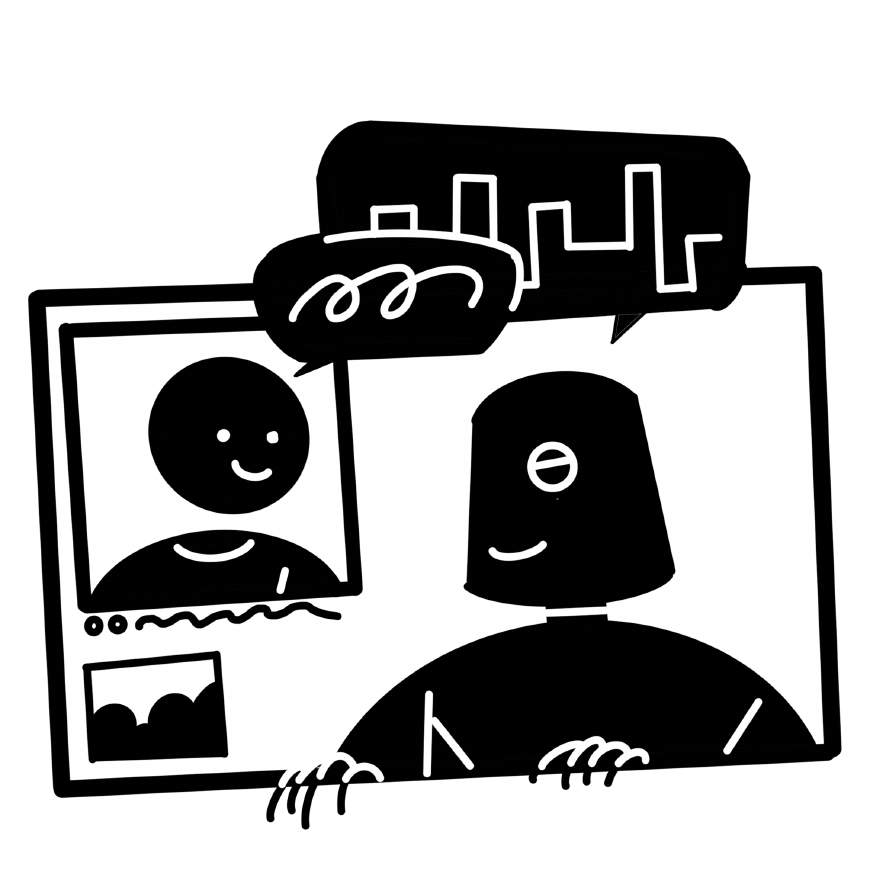}\textbf{Interaction}}: Agents are capable of interacting with other agents, the environment, and humans.
The interactive protocol can include collaboration, competition, hierarchical relationships, or specific communication principles~\cite{hua2024warpeacewaragentlarge,pan2024agentcoordvisuallyexploringcoordination}.

\item {\includegraphics[width=0.05\textwidth]{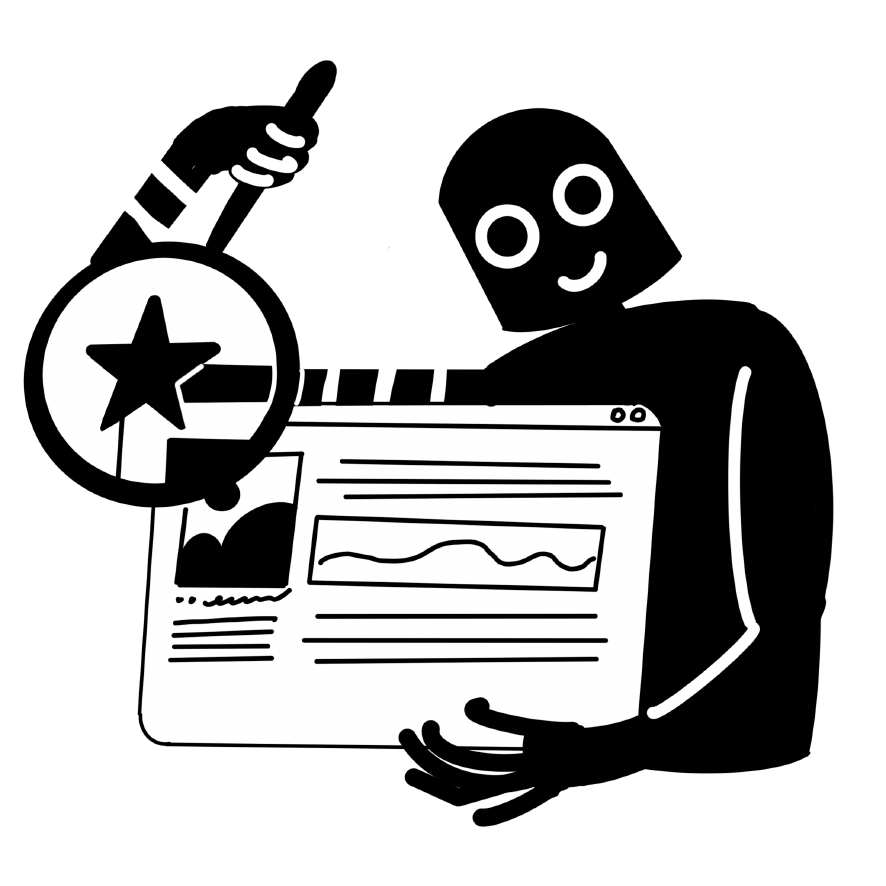}\textbf{Goal}}: Agents typically have predefined goals they strive to accomplish. 
It is worth noting that agents may have both long-term~\cite{NEURIPS2023_5950bf29, NEURIPS2023_a3621ee9} and short-term goals~\cite{pan2024agentcoordvisuallyexploringcoordination}.
Long-term goals are strategic and involve sustained effort, while short-term goals are more immediate objectives that serve as incremental steps toward achieving long-term goals~\cite{shridhar2020alfredbenchmarkinterpretinggrounded}.

\item {\includegraphics[width=0.05\textwidth]{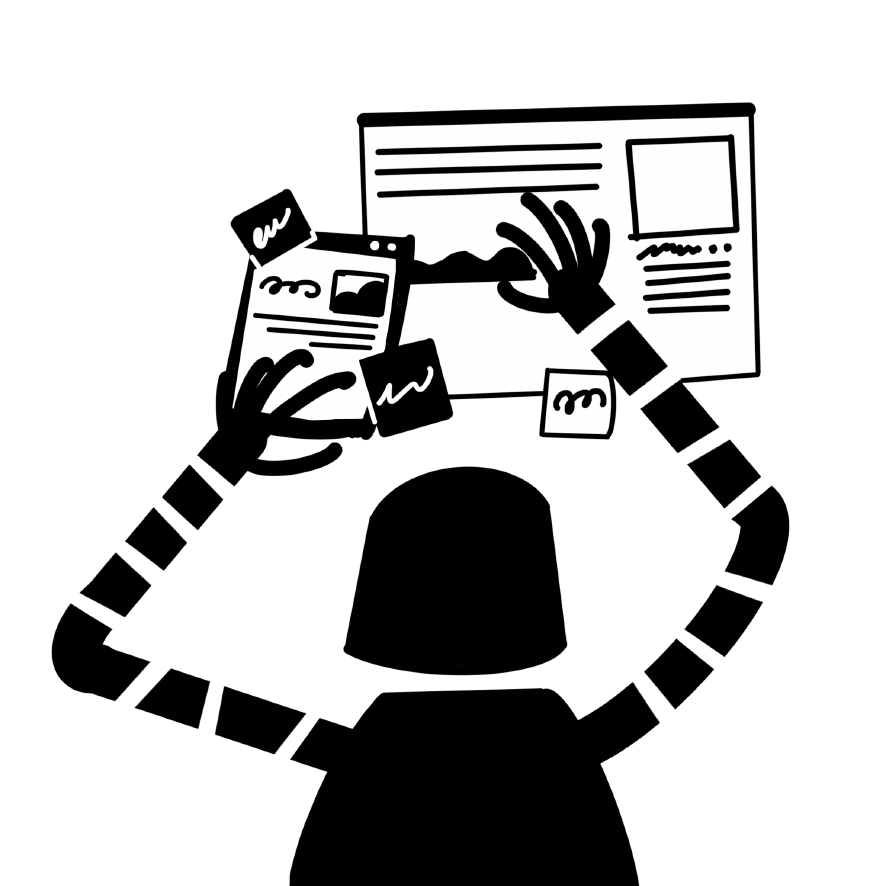}\textbf{Automony}}: Agents can function independently, making decisions and taking actions without direct human control. 
Specifically, agents adapt to environmental changes or interactions with other agents.

\item {\includegraphics[width=0.05\textwidth]{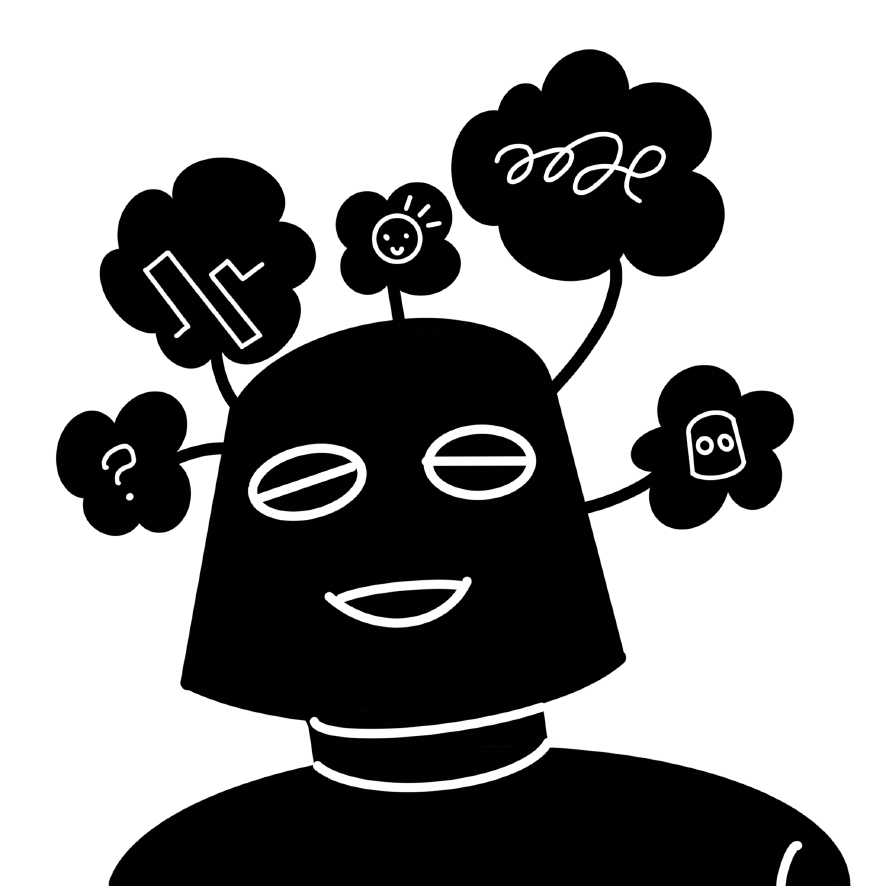}\textbf{Learning Ability}}: Agents have the capacity to learn from their experiences or adapt over time.
This learning ability enables agents to modify their behavior rules based on past outcomes or agents' memory, improving their performance or strategy as the simulation progresses~\cite{cui2024chatlawmultiagentcollaborativelegal,doi:10.1073/pnas.2115730119}.

\end{itemize}

These internal state components collectively govern the outward behaviors like humans:
\begin{itemize}
    \item {\includegraphics[width=0.05\textwidth]{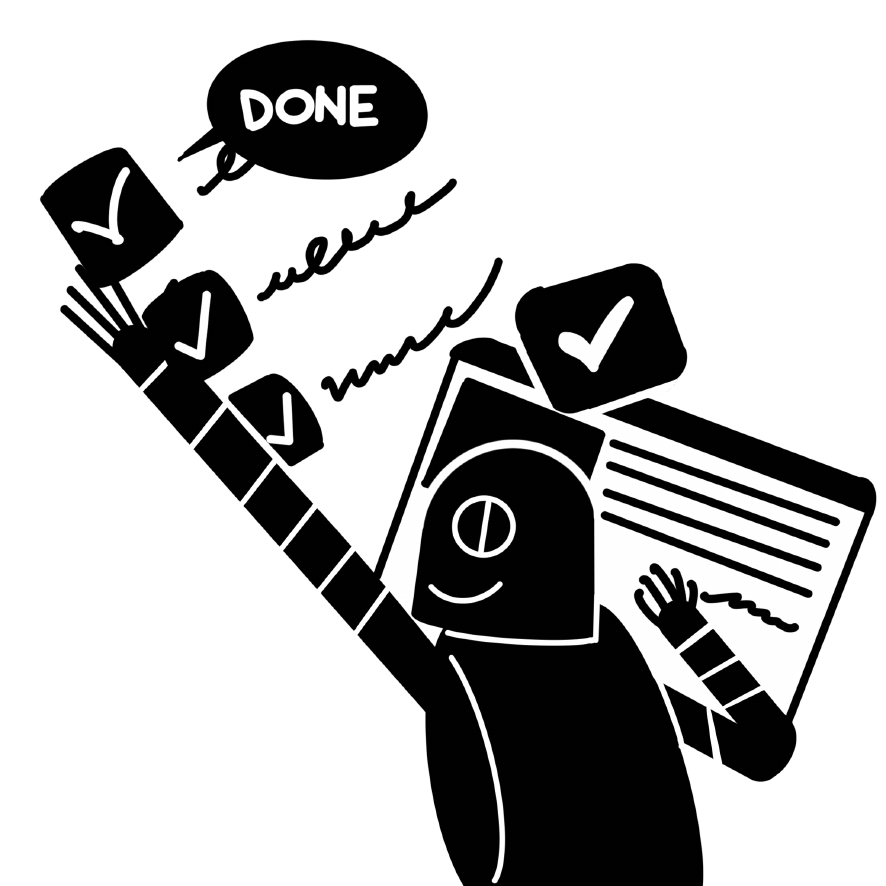}\textbf{Action}}: Observable behaviors performed by agents in response to their environment.
    These actions represent the agent's outward expression of its internal states.
\end{itemize}

Understanding both dimensions is crucial for designing realistic and effective simulations.
Together, these components allow agents to behave in human-like ways, offering rich, complex interactions that drive the sophistication of ABMS.
   
\subsubsection{Environment}
Prior to discussing the components of environments, we first present a classification of environments where agents reside.
The classification is represented across two dimensions: Physical \textit{vs.} Virtual and Real \textit{vs.} Simulated. 
This framework distinguishes environments based on their nature, either grounded in tangible, real-world settings or constructed within virtual or simulated domains.

\textit{Physical vs. Virtual}: The physical environment refers to the actual, physical world where objects, people, and places exist tangibly. 
Examples include homes, offices, streets, and natural settings.
While, the virtual environment refers to the online or digital world, which exists in cyberspace and is accessed through computers, smartphones, or other digital devices.
Examples include social media platforms, online forums, and video games.

\textit{Real vs. Simulated}:
The real environment refers to the world in which humans live and is subject to real-world laws and dynamics.
The simulated environment refers to a virtual or artificially constructed environment that mimics the dynamics of the real world or represents hypothetical scenarios.

By combining the two dimensions, four distinct quadrants are formed to help differentiate the variety of environments agents can inhabit: 
1) \textit{Real-physical} environment represents the world where humans live and interact with tangible objects.
For example, a real kitchen or a physical office where agents~(robots) perform tasks with real-world consequences~\cite{ren2023robotsaskhelpuncertainty}.
2) \textit{Simulated-physical} environment mimics artificially real-world dynamics but is not part of the tangible world.
For instance, a simulated map or virtual town layout is designed to replicate physical environments for testing or exploration purposes~\cite{Cui_2024_WACV}.
3) \textit{Real-virtual} environment is real in the sense that it reflects actual content or social contexts, but it exists in the virtual or digital realm, such as Facebook~\cite{noauthor_meta_nodate}. 
4) \textit{Simulated-virtual} environment is designed to mimic the virtual world accessed by real humans.  
For example, a virtual social media platform is constructed for simulating propagation~\cite{10.1145/3526113.3545616}.

The classification helps us understand how different types of environments are structured and define the necessary components for building effective and relevant environments.
The components of an environment encompass its fundamental structure and governing elements that shape how agents behave and interact within it:
\begin{itemize}
    \item \includegraphics[width=0.05\textwidth]{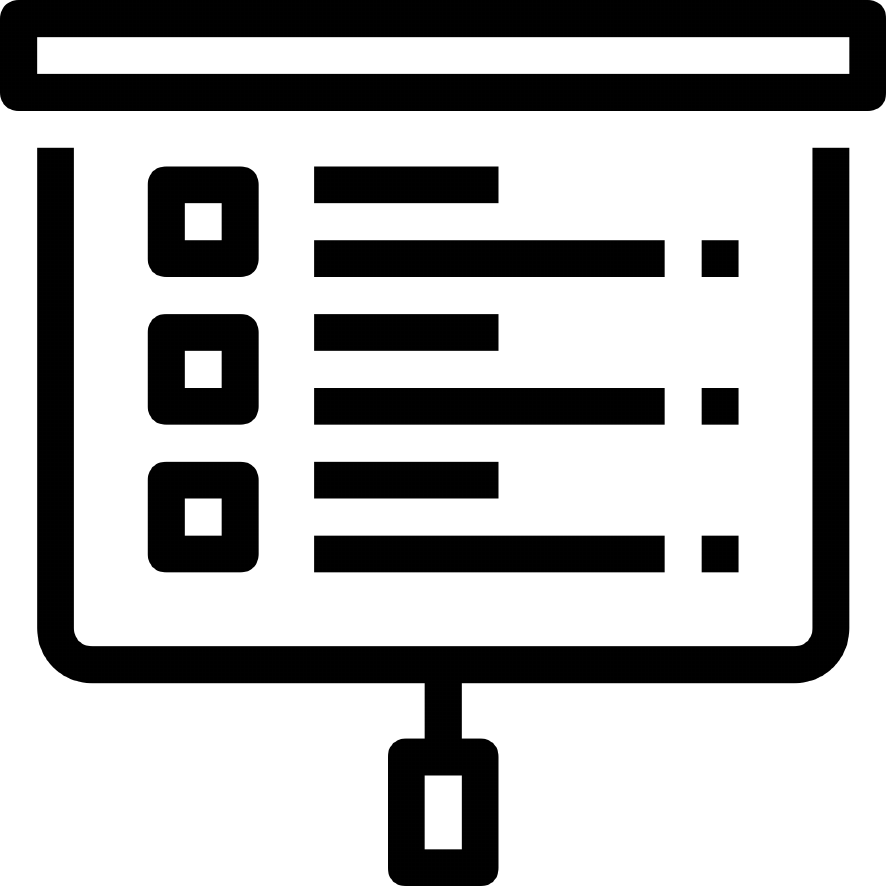}\textbf{Description}: The description of the environment outlines its key characteristics and defines the scope of the simulation or system.
    It provides a conceptual or formal representation of the environment's purpose, scale, and structure~\cite{park2023choicematessupportingunfamiliaronline,10.1145/3613904.3642159}.
    \item \includegraphics[width=0.05\textwidth]{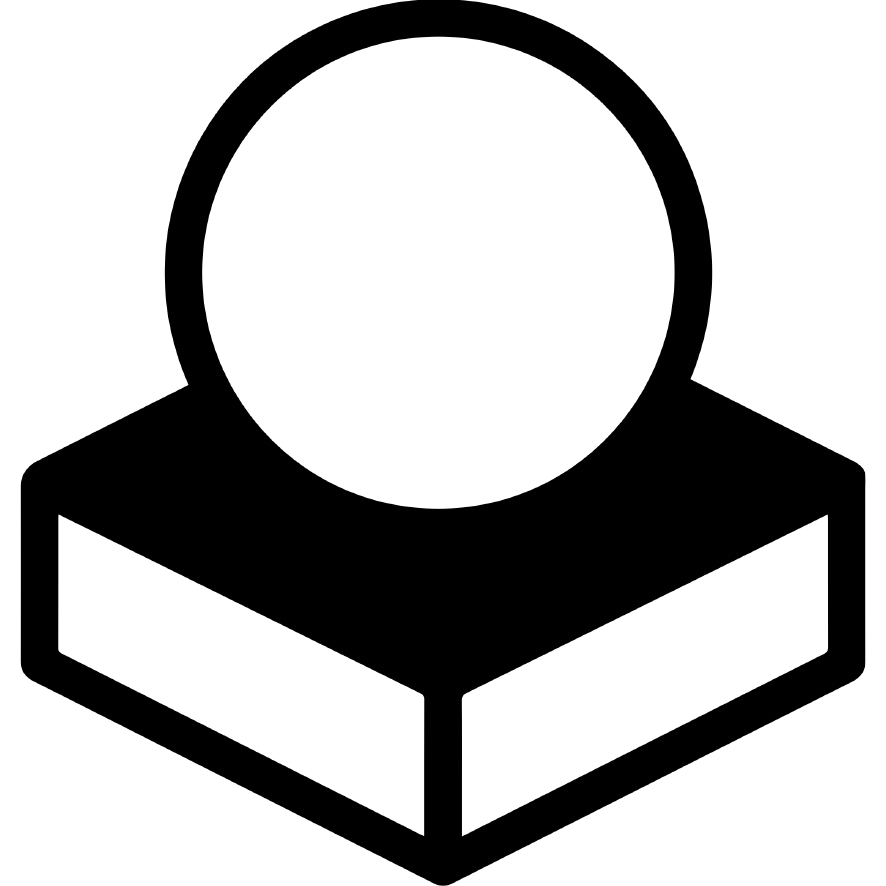}\textbf{Object}: Objects refer to the elements present within the environment with which agents can interact.
    These can include both tangible and intangible elements depending on whether the environment is physical or not.
    For example, objects may include desks or tables in a physical environment~\cite{ahn2022icanisay}.
    In a virtual environment, objects may include digital assets or virtual entities~\cite{wang2023voyageropenendedembodiedagent}.
    \item \includegraphics[width=0.05\textwidth]{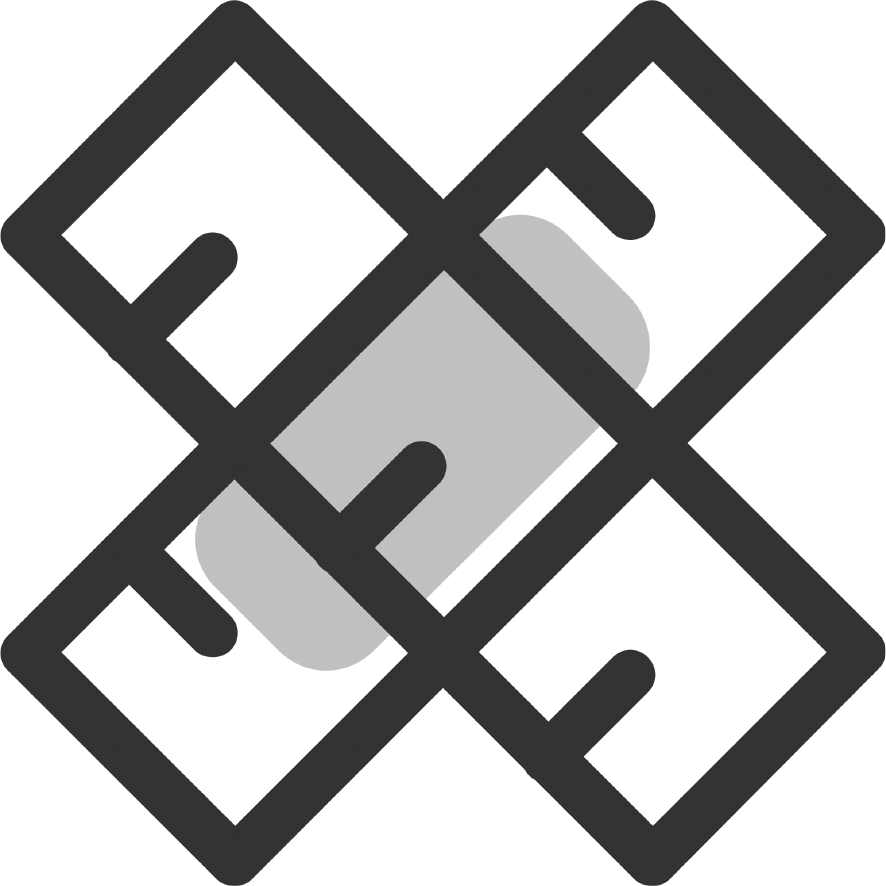}\textbf{Rule}: Rules are the foundational guidelines that dictate how agents can interact with the environment and each other. 
    They serve as the internal logic of the system, determining the possible actions agents can take and the consequences of those actions~\cite{hua2024warpeacewaragentlarge}.
    These rules often emulate real-world dynamics~(\eg gravity, economics~\cite{LENGNICK2013102}).
    Moreover, they can include limitations or incentives for specific agent behaviors, such as penalties for violating certain rules or rewards for achieving objectives~\cite{10.1145/3526113.3545616,basavatia2023complexworld}.
\end{itemize}

These three components may not all be immediately visible to agents but serve as the underlying framework of the environment.
They determine its foundational regulations, influencing how agents behave and interact at a deeper, systemic level.

\subsubsection{Simulation Configuration}
The simulation brings agents and the environment together to represent and analyze complex systems.
It tracks agents' actions, the environment's evolution, and overall system dynamics over time.
We summarize three main components of the simulation configuration:
\begin{itemize}
    \item \includegraphics[width=0.05\textwidth]{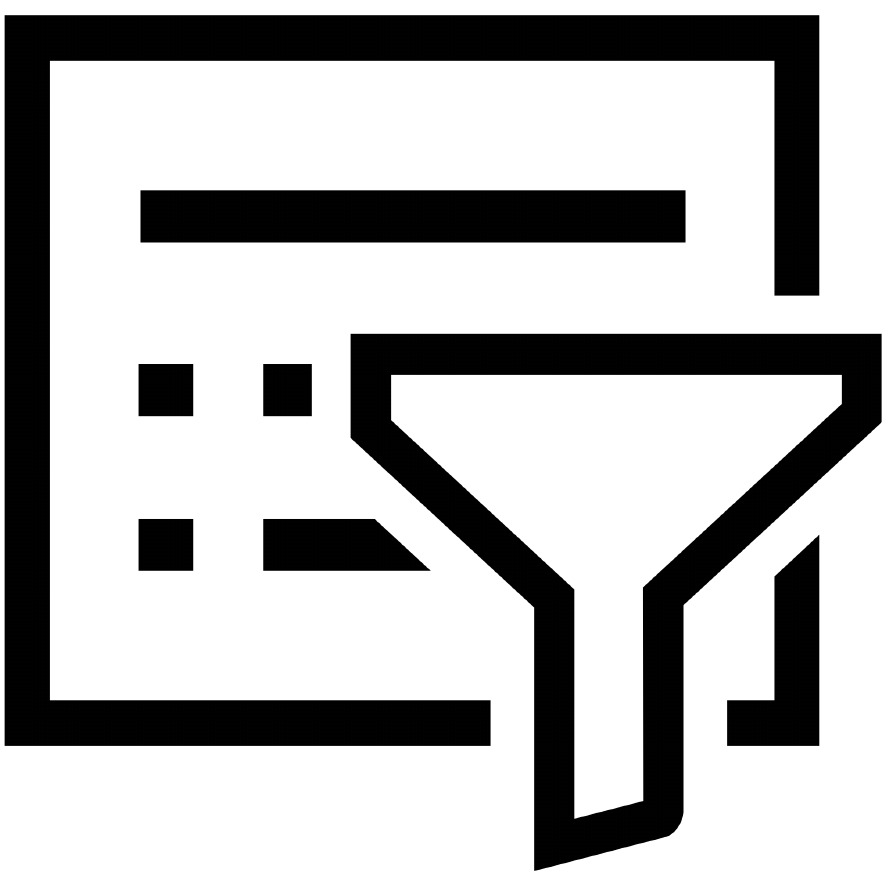}\textbf{Condition}: The running setup and parameters that define the simulation's model running state, such as the simulation's start and end time and simulation interval for discrete models.
    \item \includegraphics[width=0.05\textwidth]{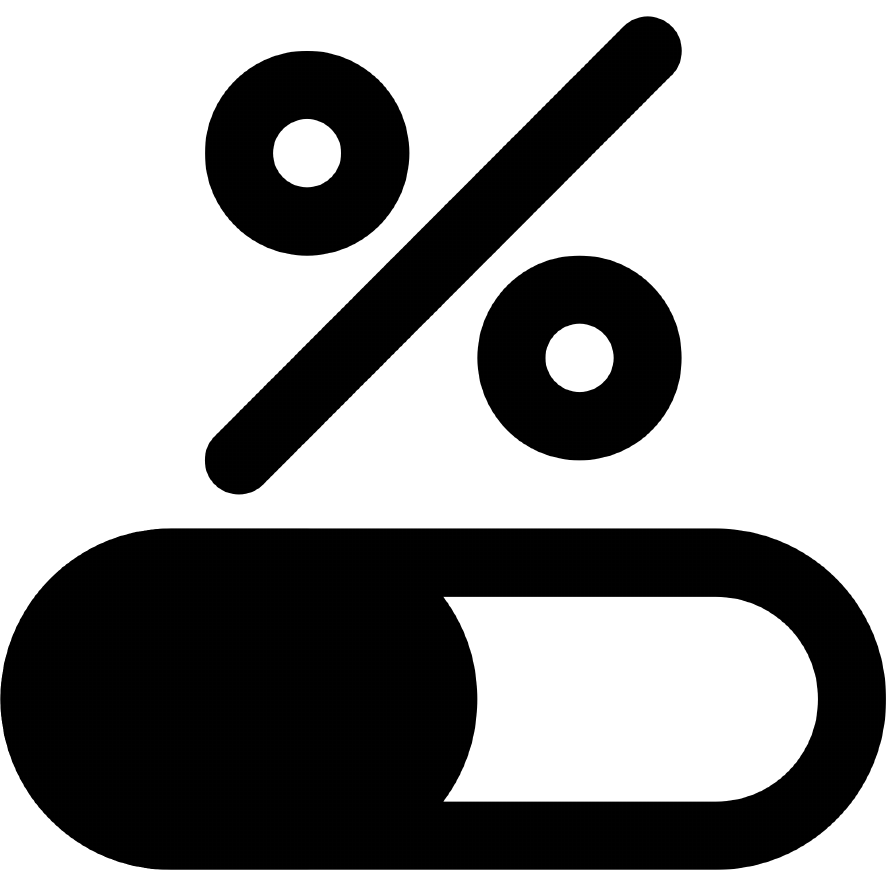}\textbf{Progress}: It tracks the temporal evolution of the simulation~\cite{chen2023agentversefacilitatingmultiagentcollaboration}.
    Agents and the environment evolve over time, and monitoring these transitions is crucial to understand the dynamics of the simulation~\cite{pan2024agentcoordvisuallyexploringcoordination}.
    \item \includegraphics[width=0.05\textwidth]{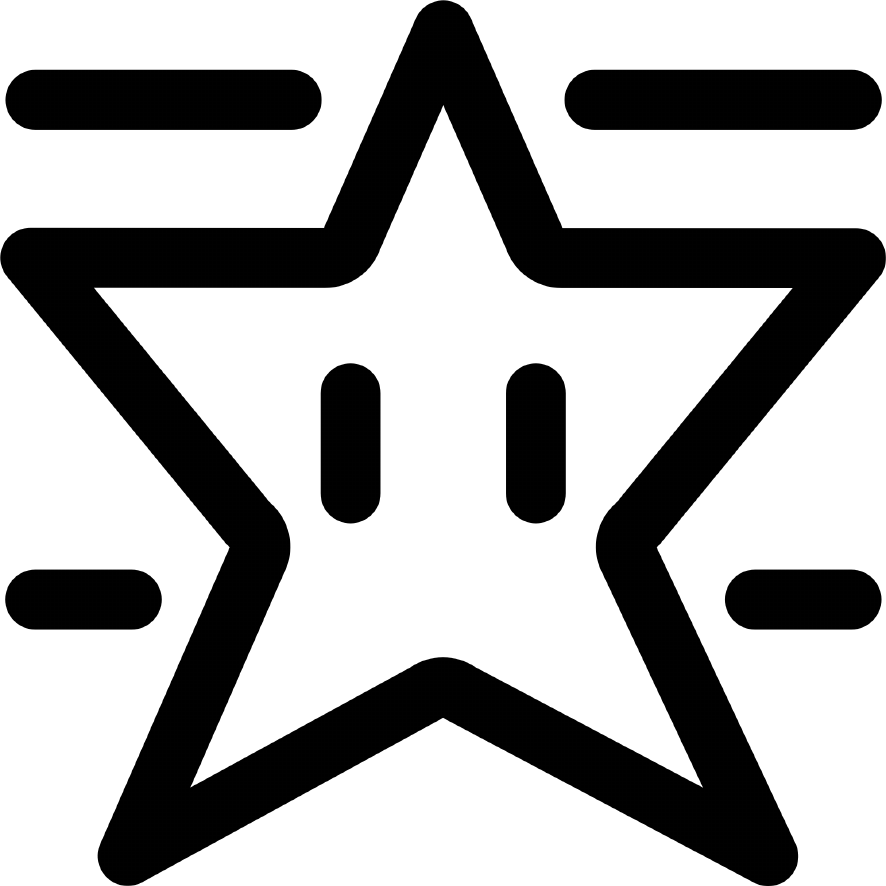}\textbf{Technique}: It refers to the computational methods and algorithms used to run the simulation.
    For example, depending on the complexity of the model, techniques such as rule-based algorithms~\cite{NEURIPS2021_86e8f7ab}, machine learning~\cite{https://doi.org/10.1111/exsy.13325}, reinforcement learning~\cite{vinyals_grandmaster_2019}, or LLMs~\cite{10.1145/3586183.3606763} may be employed to generate agent behaviors or environmental changes.
\end{itemize}

In summary, agents, environment, and simulation configuration form the three essential elements of ABMS. 
Agents act as autonomous entities within a defined environment, and their interactions and decisions are modeled through the simulation configuration, providing insights into complex systems. 

\subsection{Who: Roles of Human}
Shakespeare said, \textit{``The world is a stage and all the men and women, however, some performers, they all have off time, that the time has game.''}
We find that the roles that users play in interactions can be effectively explained through an analogy from the field of theater.
In the context of ABMS, agents can be regarded as the ``actors'' in a theatrical production, since they have predefined roles that shape their behaviors in predefined scenarios.
Therefore, we classify user roles by drawing upon professions from the theater: scriptwriter, director, actor, prototype, and observer.
It is worth noting that while these roles share similarities with those in the theater, they are not entirely identical.

\includegraphics[width=0.05\textwidth]{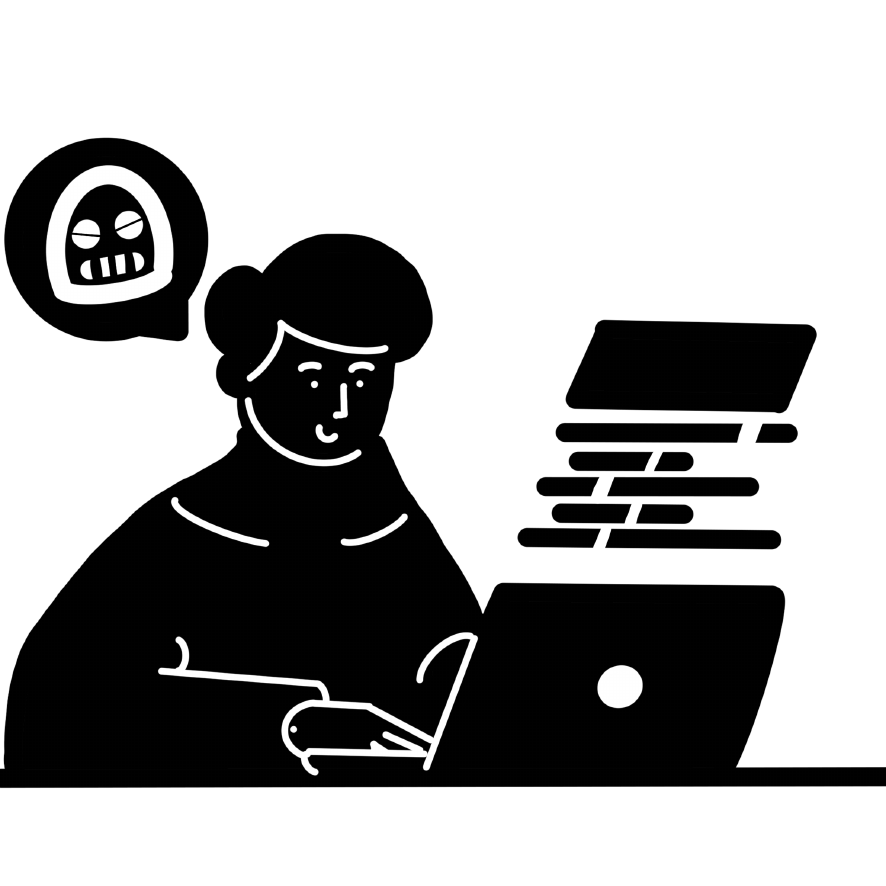}
\textit{Scriptwriter.} 
The scriptwriter initializes the purpose and structure of the simulation~\cite{10.1145/3526113.3545616,chan2023chatevalbetterllmbasedevaluators}. 
In this role, users are responsible for defining the agents and environments~\cite{lin2023agentsimsopensourcesandboxlarge}, essentially laying the foundation upon which the simulation will run.
They establish the objectives, constraints, and initial conditions to guide the simulation's progression~\cite{jinxin2023cgmiconfigurablegeneralmultiagent}.

\includegraphics[width=0.05\textwidth]{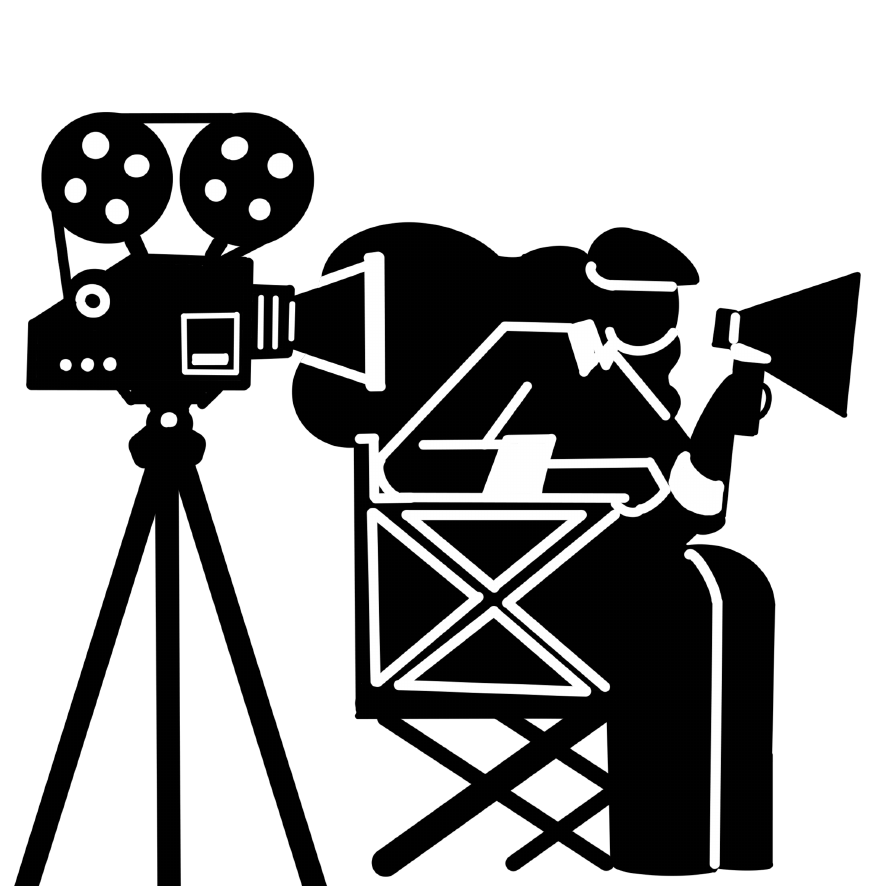}
\textit{Director.} 
As the director, the user controls the timeline and conditions for the simulation, guiding the agents and adjusting parameters during the simulation process. 
Users can direct agents, instructing them to start, pause, or restart the simulation~\cite{chen2023agentversefacilitatingmultiagentcollaboration,ren2023robotsaskhelpuncertainty}.
Users can also offer guidance to agents in a manner akin to a director instructing actors in a performance~\cite{mehta2024improvinggroundedlanguageunderstanding,unknown,park2023choicematessupportingunfamiliaronline}.
In most cases, this role emphasizes managing the flow and direction of the simulation once it is set in motion.

\includegraphics[width=0.05\textwidth]{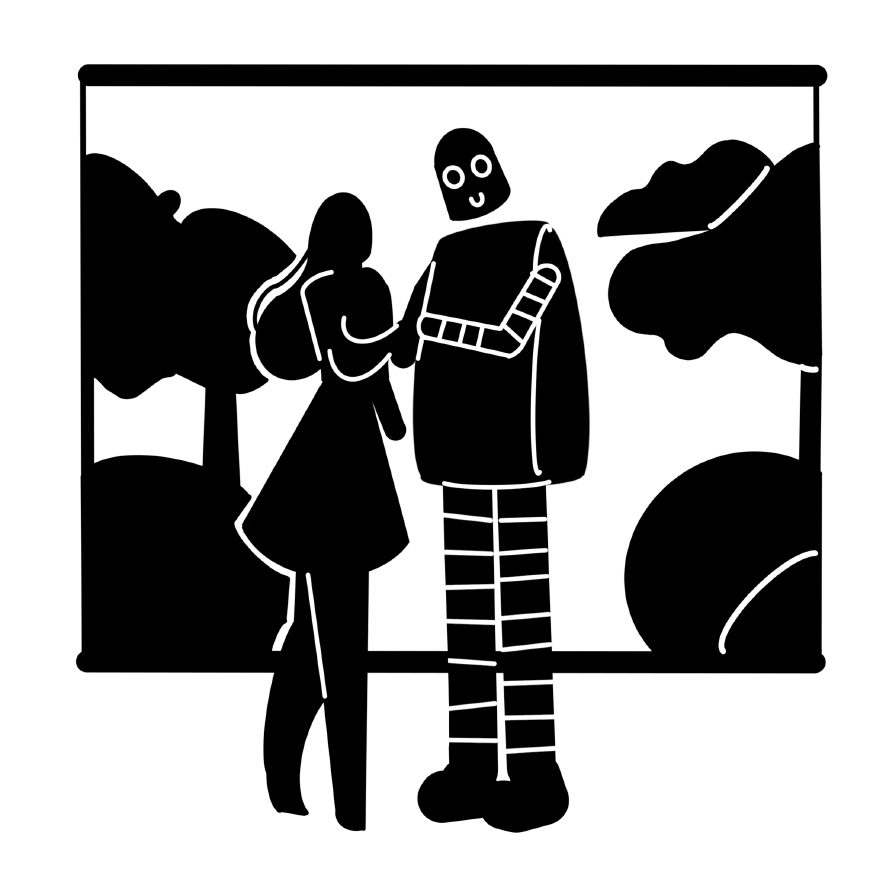}
\textit{Actor.}
The actor role represents the user interacting with the simulation as an agent, shifting from passive observation to active engagement.
Users live with other agents as if they were one of them, influencing outcomes by participating in the simulation~\cite{mao2024alympicsllmagentsmeet,zhou2024sotopiainteractiveevaluationsocial, NEURIPS2021_86e8f7ab}.
They interact with other agents or manipulate elements of the environment~\cite{10.1145/3586183.3606763,10.1145/3579598}, which can alter the course of the simulation or help achieve specific goals.
Specifically, other agents also perceive them as agents, other than humans.

\includegraphics[width=0.05\textwidth]{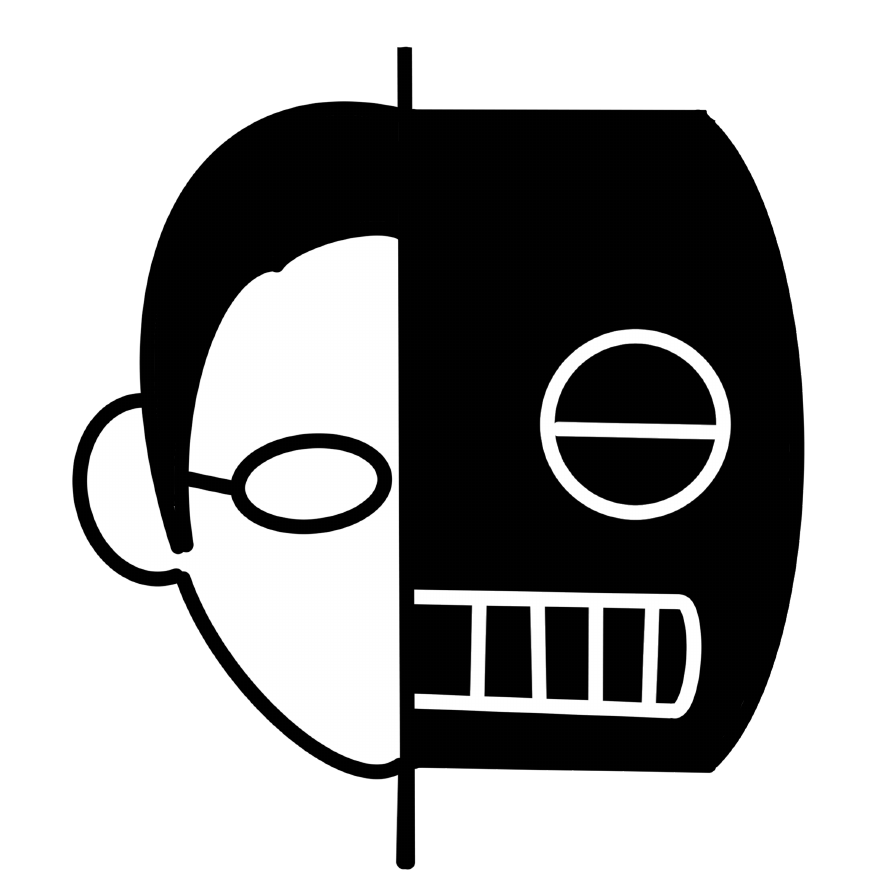}
\textit{Prototype.} 
In a theatrical context, some roles are often based on real individuals as prototypes.
Similarly, users can serve as prototypes or references for the agents within the simulation~\cite{Argyle_Busby_Fulda_Gubler_Rytting_Wingate_2023, pmlr-v202-aher23a}.
They provide a basis upon which agents' characteristics can be built.
Unlike the actor, the prototype does not directly participate in the simulation, but influences how agents are designed or programmed.

\includegraphics[width=0.05\textwidth]{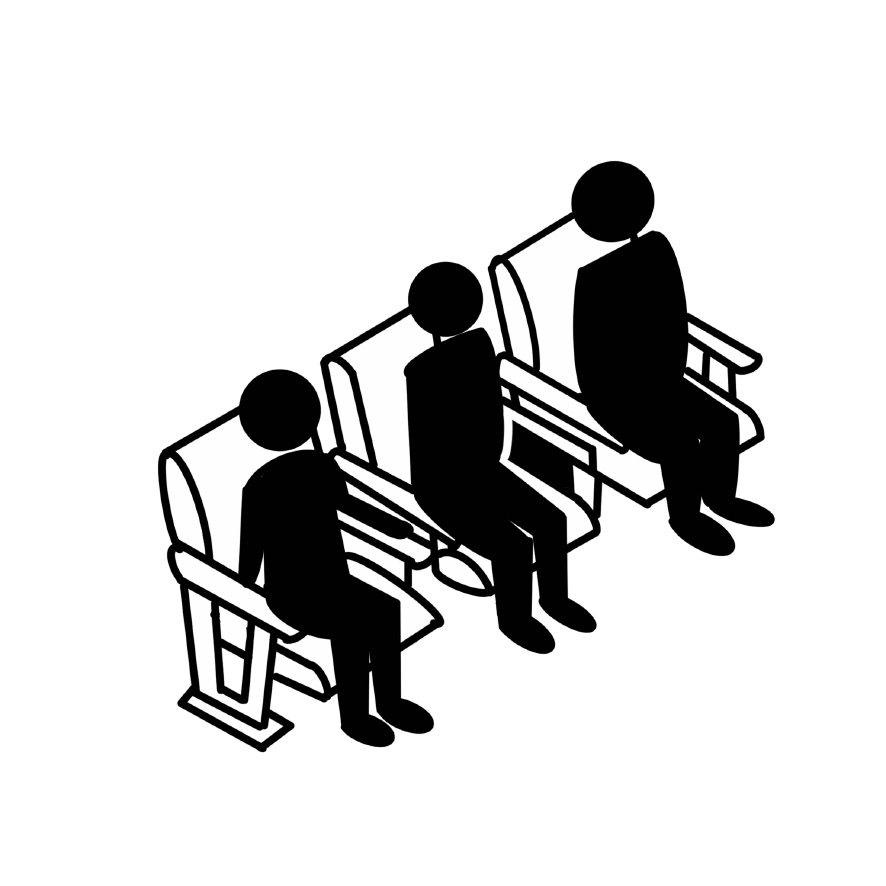}
\textit{Observer.}
The observer takes a passive yet crucial role by monitoring the simulation in real-time, gathering data and insights for further analysis~\cite{NEURIPS2023_a3621ee9,10.1145/3544548.3580688}.
Users watch the simulation unfold without intervening in the process as the audience in a theater.
They further analyze and interpret the behaviors of agents within the simulation, seeking to understand the underlying patterns, trends, or outcomes~\cite{10520238,electronics12122722,hua2024warpeacewaragentlarge}.

The environment serves as the stage where all actions occur and emergent behaviors are like the unscripted moments in a live performance. 
These five roles represent different types of user involvement with ABMS. 
Each role has a unique contribution, from defining and designing the simulation’s framework to actively participating in or passively observing its outcomes, which illustrates the flexibility and depth of user involvement in interactive simulations.

\subsection{How: Means of Interaction}
Various means of interaction allow users to engage with ABMS and ensure that users can effectively exert influence over the simulation.
The primary interaction means are as follows:

\includegraphics[width=0.05\textwidth]{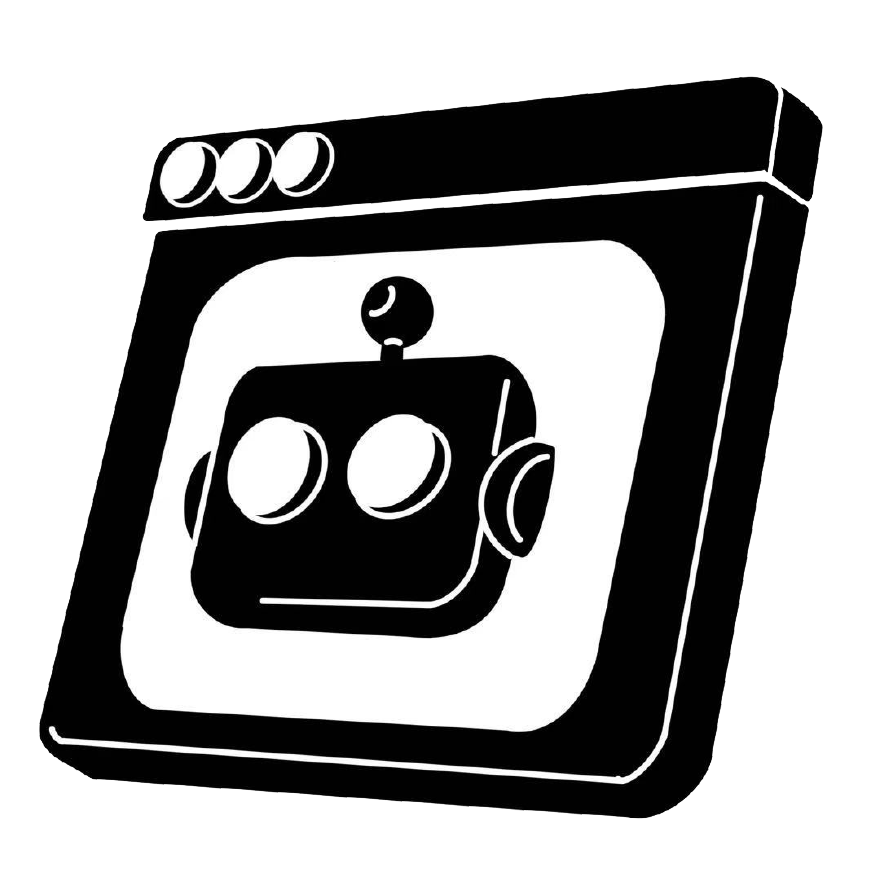}
\textit{Interface.}
The user interface~(UI) provides users access to manage the simulation.
Through buttons~\cite{chen2023agentversefacilitatingmultiagentcollaboration} and control panels~\cite{kovač2023socialaischoolinsightsdevelopmental}, users can customize various aspects of the simulation.
Graphical design~\cite{lin2023agentsimsopensourcesandboxlarge} and visualizations~\cite{pan2024agentcoordvisuallyexploringcoordination}, such as charts~\cite{10520238} and real-time agent movements~\cite{10.1145/3613904.3642159} within the environment, enable users to track agent interactions, observe emergent behaviors, and analyze the outcomes of different scenarios.
The interface often provides real-time feedback~\cite{chan2023chatevalbetterllmbasedevaluators} based on user inputs, displaying how changes in parameters affect agent behaviors and simulation outcomes.
Furthermore, it provides users with an intuitive and interactive way to control and analyze simulations, facilitating deeper engagement with the simulation and enhancing the user’s ability to draw meaningful insights.

\includegraphics[width=0.05\textwidth]{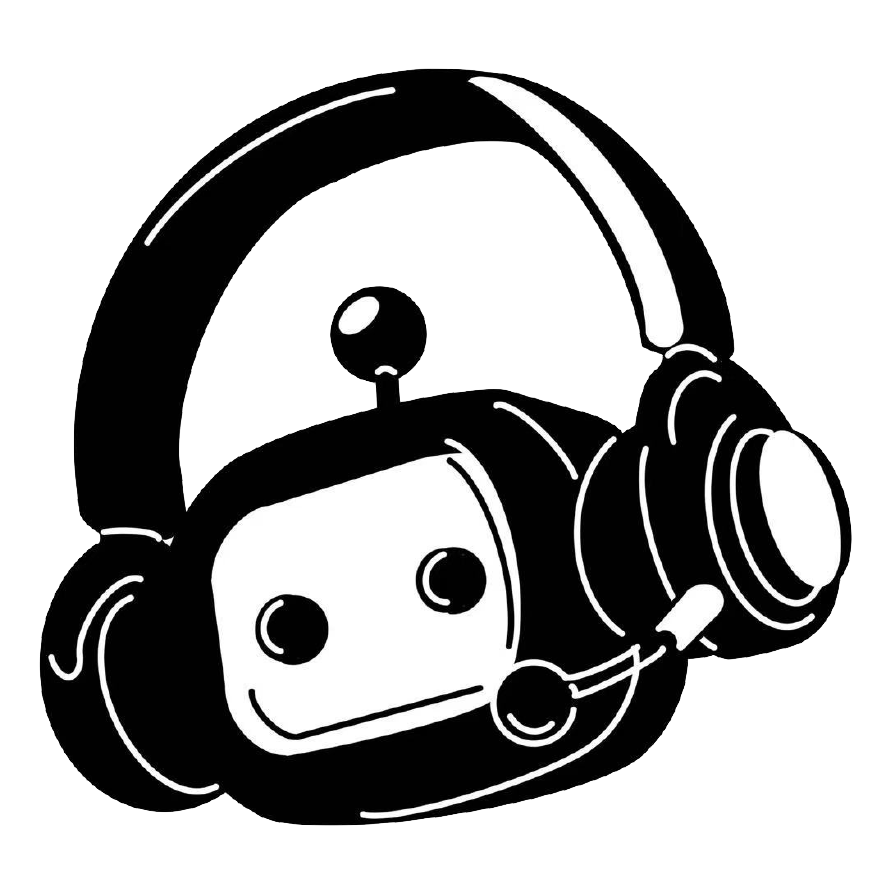}
\textit{Natural Language.}
Advances in AI and natural language processing~(NLP)~\cite{bommasani2022opportunities,brown_language_2020}, such as LLMs, enable users to give commands or ask questions in everyday language.
Users are allowed to use natural language commands to control the simulation settings, such as defining agents and environments~\cite{wang2023humanoidagentsplatformsimulating,10.1145/3526113.3545616}.
What's more, users can communicate with agents directly to guide them~\cite{shridhar2020alfredbenchmarkinterpretinggrounded} with high-level goals and low-level instructions or interview them for ``innermost thoughts''~\cite{10.1145/3586183.3606763}.

\includegraphics[width=0.05\textwidth]{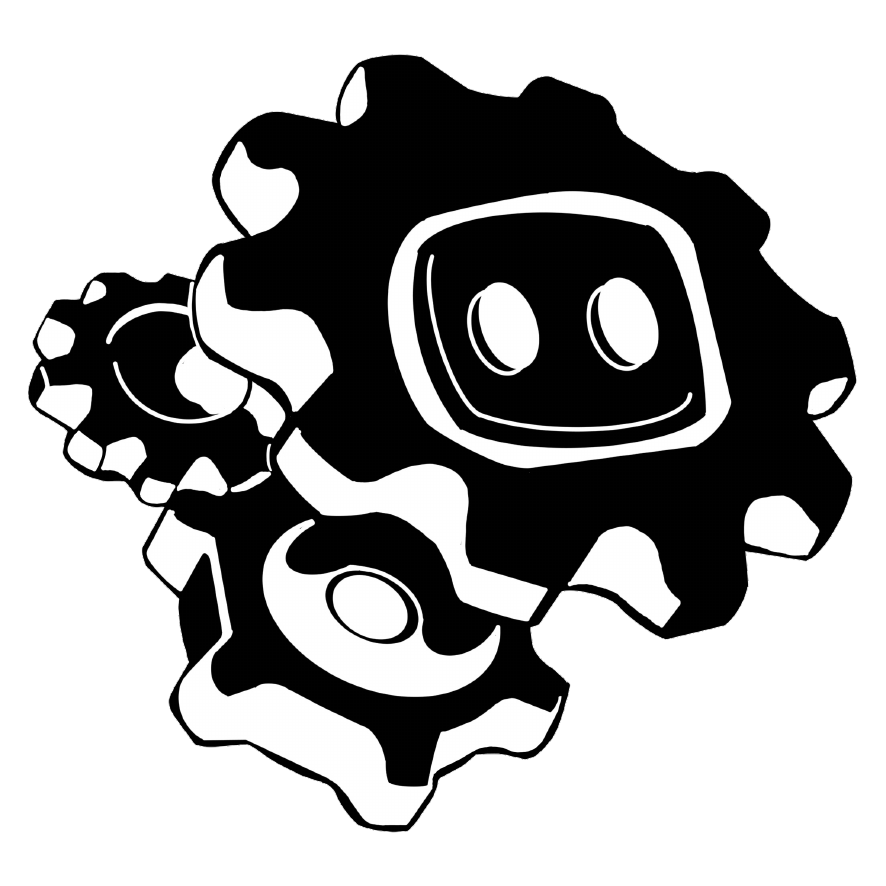}
\textit{Configuration Setting.}
We categorize methods that involve direct interaction with algorithms as configuration settings, which typically require users to have a programming background.
Configuration files~(like YAML, JSON XML) as user inputs are often used to configure simulation parameters, define agent properties, and set environmental conditions~\cite{wang2024userbehaviorsimulationlarge,hua2024warpeacewaragentlarge}.
Unlike natural language, which is flexible and often ambiguous, the structured text file follows a specific syntax and format. 
It is organized in a hierarchical or key-value structure that can be easily read and interpreted by machines.
Additionally, several libraries and APIs can be applied to construct ABMS~\cite{li2023modelscopeagentbuildingcustomizableagent}.

\includegraphics[width=0.05\textwidth]{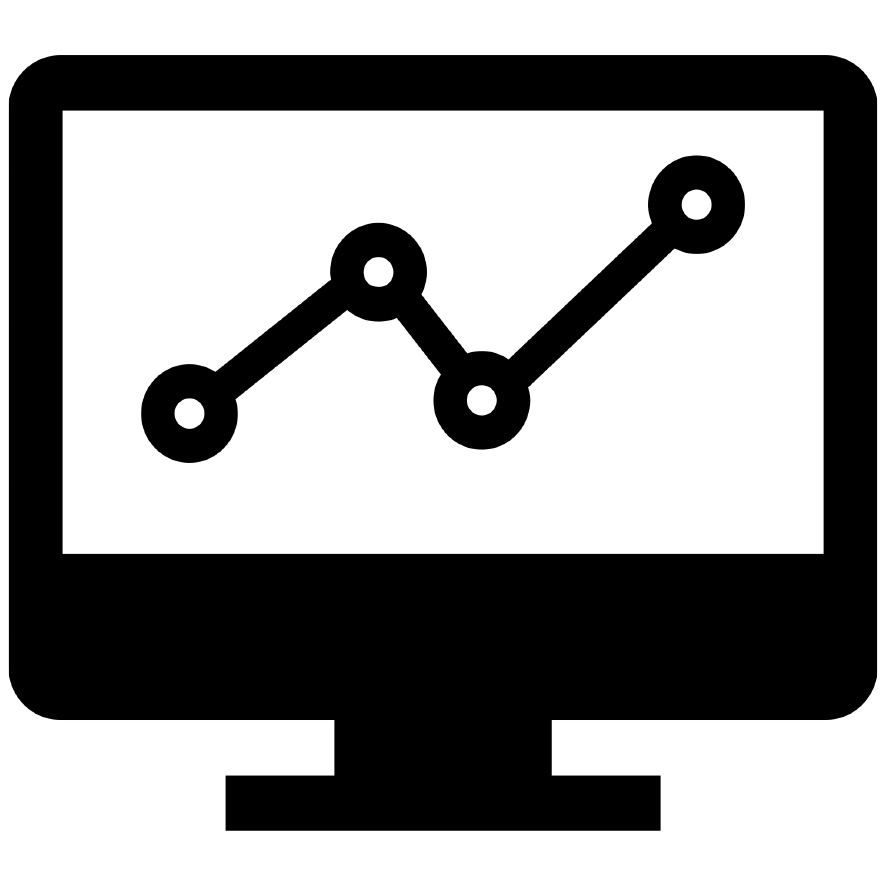}
\textit{Data Integration.}
Users can interact with ABMS with external datasets.
For example, agents can utilize users' profile data, such as demographic information, to replicate human samples for enhancing the overall realism of the simulation~\cite{Argyle_Busby_Fulda_Gubler_Rytting_Wingate_2023,gao2023s3socialnetworksimulationlarge}.
On the other hand, users gain simulation data for further analysis~\cite{10.1145/3526113.3545616}.
This data can then be analyzed to extract insights and identify patterns, allowing for informed decision-making or the refinement of the model.

\includegraphics[width=0.05\textwidth]{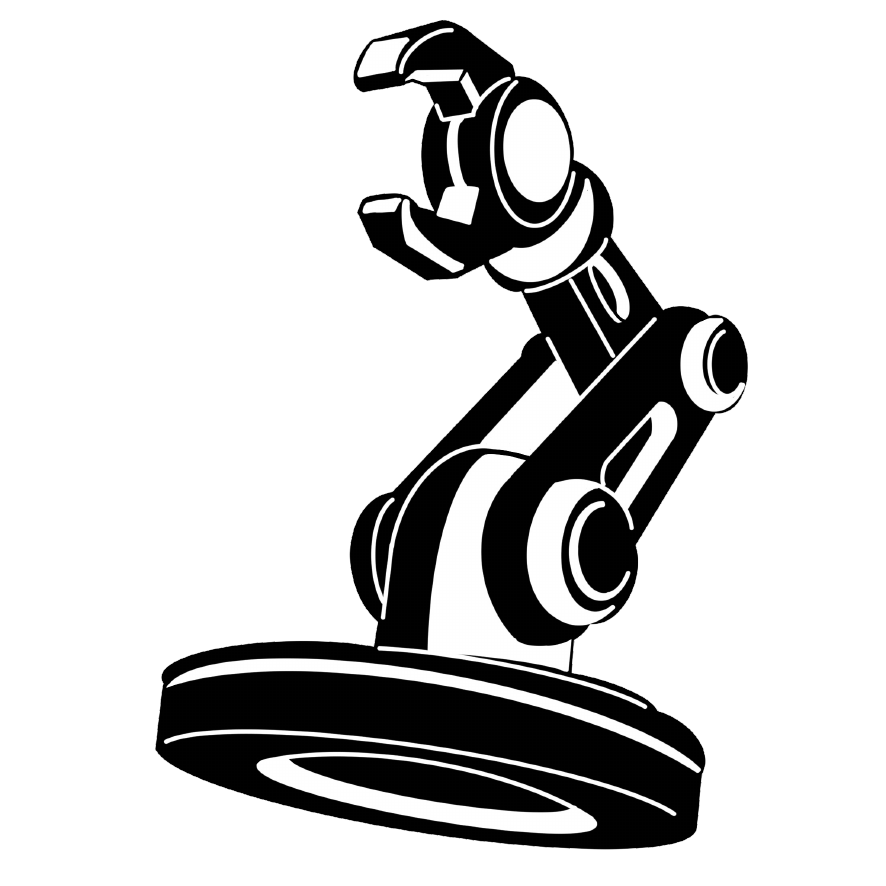}
\textit{Physical Movement.}
In certain simulations, especially those involving robotics or virtual reality/augmented reality~(VR/AR), physical movement can be a means of interaction.
Users physically interact with objects or agents in the real world, which in turn affects the simulation ~\cite{mandi2023rocodialecticmultirobotcollaboration,10.1145/3613904.3642183}.
This direct physical contact allows for real-time, hands-on control and interaction with the simulated environment.
On the other hand, in the virtual environment, users can interact with agents and the surroundings through body gestures and facial expressions~\cite{10.1145/3613905.3637145,10.1145/3613904.3642947}.

\section{Findings}\label{finding}
In this section, We demonstrate our findings organized by specific goals~(Why). 
We aim to reveal the most common human-AI interaction patterns as a focal area of study. 
Furthermore, certain patterns remain under-investigated in previous research, raising questions about their entailment and potential future applications.

\subsection{Goal 1: Initialize the Simulation}

Initializing the environment is the most frequently occurring goal in our reviewed literature.
Due to the large number of papers in this category, detailed information can be found in Appendix~\ref{Ainitial}, Table~\ref{tab:initial}.
Firstly, we observe that users interact with the models before the simulation and primarily assume three roles: scriptwriter, director, and prototype.
As scriptwriters, users need to establish a foundational background for the simulation.
Users create agents by defining their identity~\cite{hua2024warpeacewaragentlarge,lin2023agentsimsopensourcesandboxlarge}, interaction~\cite{berryman2008review}, long-term goal~\cite{hong2024metagptmetaprogrammingmultiagent}, and learning ability~\cite{li2023modelscopeagentbuildingcustomizableagent}.
Similarly, users can control the description~\cite{jinxin2023cgmiconfigurablegeneralmultiagent}, objects~\cite{basavatia2023complexworld}, and rules~\cite{10.1145/3526113.3545616} of environments.
Although some studies have utilized natural language command~\cite{hong2024metagptmetaprogrammingmultiagent} and interfaces~\cite{lin2023agentsimsopensourcesandboxlarge}, we find that a portion of the work requires users to engage in configuration settings, such as programming~\cite{netlogo}, graphical programming~\cite{Ped,doi:https://doi.org/10.1002/9781118762745.ch12}, importing packages~\cite{Significant_Gravitas_AutoGPT}, or writing configuration files~\cite{lin2023agentsimsopensourcesandboxlarge}.
Unlike interfaces and natural language commands, these methods present certain challenges for novice users when getting started.
However, they allow for a systematic, modular, and efficient setup of simulations from scratch.
How to combine the advantages of both aspects is a question worth exploring.

Another important role for the user is the director.
The director can directly issue goal commands to the model, prompting agents to begin executing the goals~\cite{rana2023sayplangroundinglargelanguage, ahn2022icanisay} or automatically trigger agents' actions through specific user actions~\cite{10.1145/3613904.3642183, arakawa2024prismobserverinterventionagenthelp}.
Additionally, the director can modify certain environmental settings during the initialization time~\cite{park2023choicematessupportingunfamiliaronline, 10.1145/3613904.3642159}.
The most commonly used means is natural language commands~\cite{10.1145/3678585}, followed by interface~\cite{pan2024agentcoordvisuallyexploringcoordination}.
Compared to the scriptwriter, the director controls the model from a more granular perspective.
Researchers can design appropriate interactions tailored to their specific research needs.
In some cases, users also appear in the role of prototypes and provide demographic data for agent identities.
Before the advancement of computational power, it was common to use sampling methods to select prototypes, and the information dimensions provided to the model were limited~\cite{GAUBE201392}.
Currently, sampling from the dataset is not necessary since the model can handle diverse, heterogeneous data directly~\cite{gao2023s3socialnetworksimulationlarge} with enhanced data processing abilities.

In this category, we can observe the evolution of simulation platforms or toolkits.
Before the maturity of NLP technologies, many works already supported users in initializing simulations.
However, these interactions were not as straightforward as natural language and involved a certain learning curve.
Initially, tools were difficult to use and challenging to learn, such as Netlogo~\cite{netlogo}, EINSTein~\cite{berryman2008review}, and MASON~\cite{doi:10.1177/0037549705058073}, which are required programming skills.
Later, tools like AnyLogic~\cite{doi:https://doi.org/10.1002/9781118762745.ch12} and PedSim~\cite{Ped} emerged, supporting graphical programming and visualizing simulation trajectories, making them more accessible and user-friendly.
With the emergence of large language models, diverse and lightweight simulation platforms have been developed~(\eg AutoGPT~\cite{Significant_Gravitas_AutoGPT} and Modelscope~\cite{li2023modelscopeagentbuildingcustomizableagent}), leveraging the interaction and generative capabilities of these models to support user-customized agents.
This advancement allows users to create tailored agent behaviors and scenarios more intuitively, expanding the flexibility and accessibility of simulation platforms.
We will further discuss the potential development of simulation platforms in Section~\ref{software}.

\subsection{Goal 2: Explore Different Scenarios}
Investigating various hypothetical scenarios enables users to examine how different assumptions or interventions might influence system dynamics.
The detailed information in this cluster is shown in Table~\ref{tab:explore}.
ChatEval~\cite{chan2023chatevalbetterllmbasedevaluators} supports multi-agent collaboration to compare only two language models' performance at once.
Thus, users need to predefine various models and conduct multiple simulations to compare the comparison results across different models.
The interactions in the remaining works occur during the simulation.

Users act directly as actors, exploring various scenarios through their own diverse behaviors, such as communicating with agents through natural language~\cite{10.1145/3586183.3606763,lin2023agentsimsopensourcesandboxlarge}.
Users can also take on the scriptwriter role, directly altering agents' foundational goals by natural language commands~\cite{10.1145/3586183.3606763}.
This type of work is relatively rare, possibly because users typically focus on exploring the impact of minor changes on the overall system rather than fundamentally altering the foundational setup of agents and the environment within the simulation.
In most cases, users act as directors, controlling the simulation process~\cite{wang2023humanoidagentsplatformsimulating, pan2024agentcoordvisuallyexploringcoordination}, adjusting environmental components~\cite{hua2024warpeacewaragentlarge}, and directing the actions of agents~\cite{DBLP:journals/corr/abs-2312-11813,10.1145/3613904.3642545}, etc.
Typically, by advancing or reversing the simulation progress, users can conduct ``what-if'' analysis~\cite{cui2024chatlawmultiagentcollaborativelegal,10.1145/3526113.3545616}.
``What-if'' analysis is crucial for ABMS, as it enables users to explore the potential effects of various changes within the system.
Users can observe how hypothetical scenarios impact agent behaviors and system dynamics by manipulating specific parameters or altering conditions. 

Current research on this topic is limited based on our review, highlighting a valuable opportunity for future researchers to explore ``what-if'' analysis in human-AI interactions in ABMS.
Such research could facilitate dynamic, in-depth analysis of ABMS and support decision-making processes, advancing the practical utility and impact of ABMS in complex scenarios.
Furthermore, the advent of LLMs enables users to explore different scenarios within the model using natural language and interface.
Designing a user-friendly, voice-enabled interactive interface that allows users to act as a real-world director, complete with a walkie-talkie and monitor screens, may hold significant potential as a research topic.
Users can also take on the role of actors, directly interacting with agents through natural language or physical movement with the advancement of immersive devices. 
They can modify or create diverse scenarios based on research needs.

\begin{table}[ht]
  \caption{This table introduces works concerning \textit{Explore Different Scenarios}. For simplicity, we
shorten the classification of environments: S-P: simulated-physical, S-V: simulated-virtual, R-P: real-physical, R-V: real-virtual. We also shorten the \textit{When} dimension: Pre-S: pre-simulation, D-S: during-simulation, Post-S: post-simulation. For the ``What'' dimension, we use icons instead of text to represent the secondary classification. Subsequent tables will also use similar abbreviations. Some works provide multiple interaction methods for the same goal, such as Generative Agents~\cite{10.1145/3586183.3606763} in this table.~\includegraphics[width=0.025\textwidth]{icon/action.pdf} represents agent \textit{action} and~\includegraphics[width=0.025\textwidth]{icon/goal.pdf} represents agent \textit{goal}.}
  \label{tab:explore}
  \begin{tabular}{lllllll}
    \toprule
    \textbf{Year}&\textbf{Work} & \textbf{Env}&\textbf{When} & \textbf{Who} & \textbf{What} & \textbf{How}\\
    \midrule
    
2023&Generative Agents~\cite{10.1145/3586183.3606763} & S-P & D-S  & Actor &Agent~\includegraphics[width=0.025\textwidth]{icon/action.pdf}& Language \\

-& -& -& D-S & Scriptwriter &Agent~\includegraphics[width=0.025\textwidth]{icon/goal.pdf}& Language \\

-& -& -& D-S & Director &Env~\includegraphics[width=0.025\textwidth]{icon/object.pdf}& Language \\

2022&Social Simulacra~\cite{10.1145/3526113.3545616}& S-V & D-S  & Director &Sim~\includegraphics[width=0.025\textwidth]{icon/progress.pdf}& Interface \\

2023&ChatEval~\cite{chan2023chatevalbetterllmbasedevaluators}&None & Pre-S & Director &Env~\includegraphics[width=0.025\textwidth]{icon/description.pdf}&Interface \\

2023&AgentSims~\cite{lin2023agentsimsopensourcesandboxlarge}&S-P  & D-S  & Actor &Agent~\includegraphics[width=0.025\textwidth]{icon/action.pdf}&Language; Interface \\

2023&WarAgent~\cite{hua2024warpeacewaragentlarge}
&S-P  & D-S  & Director &Env~\includegraphics[width=0.025\textwidth]{icon/description.pdf}&Language \\

2024&AgentCoord~\cite{pan2024agentcoordvisuallyexploringcoordination}
&S-P & D-S  & Director &Agent~\includegraphics[width=0.025\textwidth]{icon/goal.pdf}; Sim~\includegraphics[width=0.025\textwidth]{icon/progress.pdf}&Interface \\

2024&Rehearsal~\cite{10.1145/3613904.3642159}
&None & D-S  & Director &Agent~\includegraphics[width=0.025\textwidth]{icon/action.pdf}&Interface \\

2023&Humanoid Agents~\cite{wang2023humanoidagentsplatformsimulating}
&S-P & D-S  & Director &Sim~\includegraphics[width=0.025\textwidth]{icon/progress.pdf}&Interface \\

2023&UGI~\cite{DBLP:journals/corr/abs-2312-11813}
&S-P & D-S  & Director &Agent~\includegraphics[width=0.025\textwidth]{icon/action.pdf}&Language \\

2024&Zhang\etal~\cite{10.1145/3613904.3642545}
&R-V & D-S  & Director &Agent~\includegraphics[width=0.025\textwidth]{icon/action.pdf}&Interface \\
2024& ChatCam~\cite{10.1145/3699731}&R-P&D-S&Director&Agent~\includegraphics[width=0.025\textwidth]{icon/action.pdf}&Language\\
2024&Cuadra\etal~\cite{10.1145/3659624}&R-P&D-S&Director&Agent~\includegraphics[width=0.025\textwidth]{icon/goal.pdf}&Language; Interface\\
2024&CrowdBot~\cite{10.1145/3659601}&R-P&D-S&Director&Agent~\includegraphics[width=0.025\textwidth]{icon/goal.pdf}&Language; Physical\\
2024&Sheshadri\etal~\cite{10.1145/3631404}&R-P&D-S&Director&Agent~\includegraphics[width=0.025\textwidth]{icon/action.pdf}&Language\\
  \bottomrule
\end{tabular}
\end{table}

\subsection{Goal 3: Refine the Model}
When the model’s performance fails to meet expectations, improving its effectiveness requires user intervention.
There are 29 papers in this cluster, and the detailed information of papers is shown in Appendix~\ref{Arefine}, Table~\ref{tab:refine}.
Before the simulation, SocialAI School~\cite{gao2023s3socialnetworksimulationlarge}, Krishna\etal~\cite{doi:10.1073/pnas.2115730119} and Surrealdriver~\cite{jin2024surrealdriverdesigningllmpoweredgenerative} guide agents to learn from external resources, such as human natural language instructions and domain expertise data, to enhance their learning ability.
Although there is limited work in this area, it presents a promising approach to refine the model, and new interactions warrant further research.
The majority of methods are implemented during the simulation process.
Some of them also focus on agents' learning abilities.
Users can teach agent human knowledge and domain expertise~\cite{unknown, 10.1145/3613904.3642349, cui2024chatlawmultiagentcollaborativelegal} and directly manipulate memory system~\cite{10.1145/3586182.3615796}.

Some work allows users to directly take on the role of actors, collaborating with agents to complete tasks by natural language~\cite{zhang2024buildingcooperativeembodiedagents} or physical movements~\cite{mandi2023rocodialecticmultirobotcollaboration}.
More frequently, users assume the role of directors, steering agent actions~\cite{mehta2024improvinggroundedlanguageunderstanding, mohanty2023transforminghumancenteredaicollaboration, Padmakumar_Thomason_Shrivastava_Lange_Narayan-Chen_Gella_Piramuthu_Tur_Hakkani-Tur_2022}, goals~\cite{huang2022innermonologueembodiedreasoning,chen2023agentversefacilitatingmultiagentcollaboration}, and interaction~\cite{park2023choicematessupportingunfamiliaronline}.
Additionally, due to the stochastic nature of LLMs, users acting as directors can control the simulation progress through the interface by regenerating outcomes if the current results are unsatisfactory, allowing for the possibility of achieving more desirable outcomes~\cite{chen2023agentversefacilitatingmultiagentcollaboration,chan2023chatevalbetterllmbasedevaluators}.
This cluster appears to overlook the impact of environmental components on refining the model.
Users can potentially reduce obstacles for agents in completing tasks by controlling environmental components.
In addition to agents' learning abilities, users may consider enhancing agents' autonomy—an often-overlooked component in interaction design.

The design of human-AI interactions that harness the strengths of both humans and AI, enabling complementary collaboration, represents a significant area for exploration. 
This approach raises important questions about how best to structure interactions to optimize collaboration and achieve desired outcomes.
From our corpus of papers, we conclude that humans excel in creative thinking, domain expertise, and problem-solving in ambiguous situations, making them adept at tasks requiring abstract thought or out-of-the-box solutions~\cite{ren2023robotsaskhelpuncertainty, 10.1145/2282338.2282384}.
AI operates with consistent accuracy and efficiency, reducing the risk of human error and performing repetitive tasks without fatigue~\cite{10.1145/3672539.3686351}.
Combining these strengths, human-AI interaction has the potential to achieve more comprehensive outcomes, with humans providing complex reasoning abilities and AI enhancing efficiency and scalability.

\subsection{Goal 4: Evaluation the Performance}
Evaluating the ABMS's performance relies on assessing how well the simulation meets predefined goals. Human involvement is central to this process.
In this cluster, we extracted 47 interactions from 41 works.
Due to the large number of papers in this category, detailed information can be found in Appendix~\ref{Aevaluate}, Table~\ref{tab:evaluate}.
For users pre-simulation engaging with the model, the objective is to manipulate specific conditions to assess whether the outcomes align with their expectations.
For example, users can copy community rules and goals from real-world social platforms to the environment of ABMS~\cite{10.1145/3526113.3545616} or design agents' identity modeled on real-world demographic information~\cite{10.1145/3394486.3412862, Argyle_Busby_Fulda_Gubler_Rytting_Wingate_2023}.
Assessing the indistinguishability between agent actions and real user actions provides a measure of the reliability of ABMS simulation results.

Users primarily assume three roles during the simulation: director, actor, and observer.
The director assesses whether agents can adapt flexibly and effectively to the environment by assigning different goals to agents~\cite{10.1145/3643505} or intervening in agent actions~\cite{10.1145/3610170, 10.1145/3613905.3651026}.
The actor role is similar to the director, but they interact directly with agents within the environment, which allows for real-time engagement and firsthand observation of agent actions.
They test the agents' abilities in collaboration~\cite{zhang2024buildingcooperativeembodiedagents}, social interaction~\cite{zhou2024sotopiainteractiveevaluationsocial}, teaching~\cite{saha2023languagemodelsteachweaker}, and strategic gameplay~\cite{NEURIPS2021_86e8f7ab,10.1145/3613905.3650853}.
The observer evaluates agents by tracking their behaviors through graphical interfaces~\cite{park2023choicematessupportingunfamiliaronline,lin2023agentsimsopensourcesandboxlarge,wang2023humanoidagentsplatformsimulating} or log data~\cite{babyagi}, allowing for a detailed assessment of agent actions.

After the simulation, the majority of users, acting as observers, evaluate model performance primarily through analyzing agent action data.
They assess whether the agents perform effectively~\cite{hua2024warpeacewaragentlarge,park2023choicematessupportingunfamiliaronline} or exhibit noticeable differences from real human behaviors~\cite{10.1145/3544548.3580688,10.1145/3526113.3545616,https://doi.org/10.1111/mila.12466}.
MetaGPT~\cite{hong2024metagptmetaprogrammingmultiagent} and BactoWars~\cite{berryman2008review} provide users with interactive interfaces and videos to showcase agent performances.
Notably, Generative Agents~\cite{10.1145/3586183.3606763} proposed a unique evaluation method, interviewing agents as an actor ``reporter''.
After ``two-day'' simulated lives, by designing targeted questions, users can assess whether the agent has self-awareness of its identity, accurate memory, and action aligned with its assigned character traits.
Previous studies have largely overlooked agents' internal states. 
Future research could benefit from emphasizing the alignment between agents' internal states and outward behaviors.

LLM-powered agents are capable of simulating various human-like behaviors and reflecting different characteristics.
The coherence and consistency of LLMs' outputs make agents' behaviors more realistic and believable.
When assessing the believability of simulated behaviors, simplistic quantitative statistical methods are often inadequate. 
In these instances, human qualitative evaluations, such as the Turing test~\cite{Turing2009}, are frequently employed in research to provide more nuanced insights.
It suggests that the advent of LLMs not only introduces new interactions for users in ABMS, but also creates additional interaction requirements.
Designing reliable user experiments to evaluate agent-human resemblance presents several challenges.
Key issues include minimizing user subjectivity to prevent it from skewing evaluation results and determining whether agent behavior alone can reliably indicate human likeness.
Another complexity is interpreting agents' unusual or seemingly illogical actions; while such behaviors might suggest limitations in the agent's mimicry ability, human behavior itself often includes an element of randomness.

\subsection{Goal 5: Analyze Simulation Data}
Analyzing data generated from the ABMS process is a key goal for users.
The datasets involve logs of agents' actions, records of agents' internal state, formation and evolution of networks among agents, spatial and temporal data, etc.
By analyzing the data, users gain insights into system dynamics to support the decision-making process ultimately.
The detailed information about the literature is shown in Table~\ref{tab:analyze}.
Users all act as observers to analyze agents' actions through the interface.
In contrast to assessing the model itself, users analyze data to derive insights for downstream tasks, such as informing real-world decision-making or enhancing predictive capabilities.
Out of the ten works, six are early-developed simulation platforms or toolkits. 
This type of more mature toolkit typically provides users with data analysis modules.
EINSTein, MANA~\cite{berryman2008review}, and Swarm~\cite{minar1996swarm} display basic statistical metrics and visualization, such as tallies of agents detected and killed in battlefield and a time series graph of population dynamics.
Humanoid Agents~\cite{wang2023humanoidagentsplatformsimulating} and AnyLogic~\cite{doi:https://doi.org/10.1002/9781118762745.ch12} both provide a dashboard for users to explore agents' actions over time interactively.
Furthermore, AgentLens~\cite{10520238} and AgentCoord~\cite{cui2024chatlawmultiagentcollaborativelegal} proposed more intricate visual analytics systems to support users interactively investigating details and causes of agents' actions and multi-agent interaction strategy.
We find that the data has evolved from simple statistical metrics to complex, multi-dimensional, heterogeneous forms, such as agent emotions, diverse actions and locations, and dynamic social networks.

The integration of LLMs significantly enhances the richness and complexity of simulation data, which introduces challenges in managing, processing, and interpreting the increased intricacy of the data.
Correspondingly, the evolution of analytical tools, from basic statistical charts to dashboards and then to fully integrated visual analytics systems, reveals an increase in both their analytical capabilities and level of interactivity.
They support more nuanced insights, facilitate decision-making, and allow users to engage with complex data landscapes in a more intuitive, interactive manner.
The development of effective and efficient tools suited for analyzing ABMS data holds substantial potential research value.
For example, integrating machine learning models for data regression or classification could be considered, as well as incorporating NLP techniques to allow users to control the analysis process through natural language commands.
Regarding the \textit{When} dimension, we discover that only a limited number of works support real-time data analysis by users~(during-simulation).
Currently, real-time data analysis is challenging to implement, especially for ABMS developed with LLMs, as they can lead to unstable data generation and low processing efficiency.
Developing stable, real-time, and user-friendly analytical tools requires further investigation.

\begin{table}[ht]
  \caption{This table introduces works concerning \textit{Analyze Simulation Data}.~\includegraphics[width=0.025\textwidth]{icon/action.pdf} represents \textit{action}. }
  \label{tab:analyze}
  \begin{tabular}{lllllll}
    \toprule
    \textbf{Year}&\textbf{Work} & \textbf{Env}&\textbf{When} & \textbf{Who} & \textbf{What} & \textbf{How}\\
    \midrule
-& AnyLogic~\cite{doi:https://doi.org/10.1002/9781118762745.ch12} & S-P & D-S    & Observer  & Agent~\includegraphics[width=0.025\textwidth]{icon/action.pdf} & Interface   \\
-& EINSTein~\cite{berryman2008review} & S-P & Post-S   & Observer  & Agent~\includegraphics[width=0.025\textwidth]{icon/action.pdf} & Interface   \\
2006 & MANA~\cite{berryman2008review} & S-P & Post-S   & Observer  & Agent~\includegraphics[width=0.025\textwidth]{icon/action.pdf} & Interface   \\
1999 & NetLogo~\cite{netlogo} & S-P & Post-S   & Observer  & Agent~\includegraphics[width=0.025\textwidth]{icon/action.pdf} & Interface  \\
2006 & North\etal~\cite{10.1145/1122012.1122013} & S-P & Post-S   & Observer  & Agent~\includegraphics[width=0.025\textwidth]{icon/action.pdf} & Interface   \\
1996 & Swarm~\cite{minar1996swarm} & S-P & D-S    & Observer  & Agent~\includegraphics[width=0.025\textwidth]{icon/action.pdf} & Interface   \\
2023 & Zarzà\etal~\cite{electronics12122722} & S-P & Post-S   & Observer  & Agent~\includegraphics[width=0.025\textwidth]{icon/action.pdf} & Interface   \\
2024 & AgentLens~\cite{10520238} & S-P & Post-S   & Observer  & Agent~\includegraphics[width=0.025\textwidth]{icon/action.pdf} & Interface   \\
2024 & AgentCoord~\cite{pan2024agentcoordvisuallyexploringcoordination} & S-P & D-S    & Observer  & Agent~\includegraphics[width=0.025\textwidth]{icon/action.pdf}~\includegraphics[width=0.025\textwidth]{icon/interaction.pdf} & Interface   \\
2023 & Humanoid Agents~\cite{wang2023humanoidagentsplatformsimulating} & S-P & Post-S   & Observer  & Agent~\includegraphics[width=0.025\textwidth]{icon/action.pdf} & Interface   \\

  \bottomrule
\end{tabular}
\end{table}


\subsection{Goal 6: Be Immersed in the Environment}\label{immersed}

Immersion in the environment highlights the user’s experience within the simulation, primarily emphasizing engagement rather than control or modification.
The number of papers in this category is relatively small compared to other categories.
According to Table~\ref{tab:immerse}, there are only two papers in the category: Generative Agents~\cite{10.1145/3586183.3606763} and Alympics~\cite{mao2024alympicsllmagentsmeet}.
In both works, users can play as actors and interact with agents as if they were one of them during the simulation in the environment.
In Generative Agents, users can communicate with agents as ``mayor'' or ``reporter'' and change the status of surrounding objects.
In Alympics, human players are engaged in the game with agent players.
The user does not have a predetermined goal but seeks immersion and emotional value in the interaction process in both cases.
Due to the limited work in this area, many interactions remain to be developed.
Users can take on the role of scriptwriter or director,  granting them the ability to control the model from a ``god’s-eye view'' and effectively orchestrate the entire simulation.
This high-level perspective fosters a strong sense of engagement and immersion as users can actively influence the model's narrative and dynamics.
Besides, immersive experience in the virtual reality environment constitutes a significant and valuable area of research.
Users can interact with agents through physical movements and natural language, creating a more intuitive engagement.
We provide a further discussion on immersive experience in Sections~\ref{immersive}.

\begin{table}[ht]
  \caption{This table introduces works concerning \textit{Be Immersed in the Environment}. ~\includegraphics[width=0.025\textwidth]{icon/object.pdf} represents \textit{object}.}
  \label{tab:immerse}
  \begin{tabular}{lllllll}
    \toprule
    \textbf{Year}&\textbf{Work} & \textbf{Env}&\textbf{When} & \textbf{Who} & \textbf{What} & \textbf{How}\\
    \midrule
    2023&Generative Agents~\cite{10.1145/3586183.3606763} & S-P & D-S  & Actor &Agent~\includegraphics[width=0.025\textwidth]{icon/action.pdf};
Env~\includegraphics[width=0.025\textwidth]{icon/object.pdf}& Language\\
2023&Alympics~\cite{mao2024alympicsllmagentsmeet}&S-V & D-S  & Actor &Agent~\includegraphics[width=0.025\textwidth]{icon/action.pdf}&Interface\\
  \bottomrule
\end{tabular}
\end{table}

\subsection{Application of the Taxonomy}
Our taxonomy and findings can be used in designing human-AI interactions in ABMS that support users' customized implementation to meet research needs.
First, identify the primary goal for interaction~(\textit{Why}). 
We have summarized six goals in~\Cref{goal} that require human involvement to achieve.
Designers determine interaction goals based on our framework to address the practical needs of different research tasks.
According to the goal, designers can find existing interactions in~\Cref{finding}, including the other four dimensions~(\textit{When}, \textit{What}, \textit{Who}, and \textit{How}).
Designers can select the most appropriate interaction from the patterns or be inspired by the potential interactions we have summarized.
Designers must comprehensively consider many aspects to determine the four dimensions, including further refining interaction goals, the feasibility of technical implementation, and other relevant factors.
\section{Suggestions and Research Opportunities}
In this section, we present specific research opportunities identified through the findings in~\Cref{finding} using the proposed taxonomy in~\Cref{framework}.

\subsection{Maxmize the Potential of LLMs}
LLMs are becoming increasingly significant in enhancing interaction in ABMS due to their unique ability to understand and generate intricate human language.
By leveraging LLMs, users benefit from a more intuitive and effective interaction process.
Well-designed prompts can guide LLMs in better simulating human-like behaviors, producing contextually accurate responses, and performing complex tasks autonomously.
Users who lack knowledge of LLMs may struggle to phrase prompts in ways that yield the desired outcomes, which can lead to potentially confusing or unintended results.
Additionally, the complexity of ABMS can further complicate prompt formulation, as users must consider both the model's interpretive limits and the nuances of simulation parameters and agent behaviors.
For example, Generative Agents~\cite{10.1145/3586183.3606763} supports utilizing one paragraph of natural language description to define agent’s identity, including jobs and past experience.
Although such a design provides users with substantial freedom, it can lead to a dilemma where users are uncertain about what to write and may struggle to determine which information is essential to include in the prompt. 
This uncertainty can result in prompts that are either incomplete or overly detailed, diminishing the interaction's effectiveness.
Prompt engineering~\cite{giray_prompt_2023}, which is the process of carefully designing prompts to guide LLMs in generating accurate and contextually appropriate responses, has been widely developed, such as chain-of-thought~\cite{NEURIPS2022_9d560961} and tree-of-thought~\cite{NEURIPS2023_271db992} strategies.
We think the community could explore ways to help users craft effective prompts during interactions.
This research could involve developing adaptive prompt templates tailored to specific tasks, recommending contextually relevant prompts based on the user’s goals, or implementing prompt engineering techniques to refine users' inputs for better results. 
For example, when users must set an agent's identity through natural language, a template can be provided to guide users in specifying the required demographic information (\eg gender, age, occupation).
These approaches aim to reduce the learning curve associated with prompt creation, especially for users less familiar with LLMs, and improve the overall effectiveness of human-AI interactions.

An increasing number of specialized fields are utilizing interactive ABMS, with LLMs simulating various human roles or professions.
However, simply employing LLMs for basic question-and-answer interactions does not effectively simulate all roles, particularly those requiring domain expertise or complex reasoning abilities.
For roles like these, a more sophisticated approach is needed to capture the depth and nuance of their knowledge and thinking processes.
One possible future direction is to design cognitive architecture for agents to simulate the human thinking processes, such as retrieving and reflecting~\cite{10.1145/3586183.3606763}.
These architectures could enable more realistic and contextually aware responses to model complex human behaviors, making them more effective in roles requiring higher expertise and adaptive decision-making.
Instruction tuning~\cite{zhang2024instructiontuninglargelanguage} is another strategy to improve the performance of LLMs by training them to follow specific types of instructions more accurately.
By fine-tuning models with instruction data specific to a field, LLMs can better understand and execute nuanced, technically complex instructions that align with domain professionals' expectations.
Instruction tunning techniques have been applied in various domains~\cite{zhang2023multitaskinstructiontuningllama,liu2023goatfinetunedllamaoutperforms}, however, there is limited research addressing it in the HCI community. 
We hope that our research can inspire future researchers in this area.


\subsection{Simulation Software Development}\label{software}
Before the maturity of natural language technologies, users typically built ABMS on simulation software platforms~\cite{doi:10.1177/0037549706073695,berryman2008review}. 
These platforms did not support natural language interaction, requiring users to rely on more technical interfaces, which also involved a certain learning cost.
ABMS simulation platform with integrated natural language processing techniques may be required to enable users to interact with agents and control simulations using natural language commands, enhancing accessibility and ease of use. 
The platform could make ABMS more user-friendly and applicable across various domains, even for those without programming expertise.
Although there exist some platforms that enable the creation, deployment, and management of agents leveraging LLMs, such as autoGPT~\cite{Significant_Gravitas_AutoGPT}, AgentTorch~\cite{chopra2023agenttorch}, they still require users to have a certain level of programming knowledge.
It is important to design simulation software accessible to users with minimal technical expertise by incorporating natural language processing capabilities. 
In addition to implementing natural language interaction, other AI technologies could also be considered. 
For example, integrate machine learning algorithms to recommend relevant commands or next steps to users based on the user’s current actions, simulation state, or previous interaction sequences. 

ABMS is a versatile tool applied across numerous fields to simulate complex systems, analyze collective behaviors, and make predictions.
Different fields have unique design requirements for interactive ABMS platforms.
Each domain may prioritize distinct features, interaction methods, and data integration needs to meet specific goals effectively.
For example, economic simulations prioritize high-frequency interaction options, such as adjusting market parameters or agent strategies in real-time~\cite{helbing_agent-based_2012}.
While simulations in social science often need agents with complex, varied behaviors to model interactions like group dynamics, migration, or policy effects~\cite{gao2023s3socialnetworksimulationlarge}.
Developing simulation platforms for specific domains may empower professionals and researchers to address real-world challenges.
They could include agents and models prebuilt for the domain, tailor the interface and interaction options to the specific needs of the field, and offer analysis tools and visualization options that highlight metrics crucial to the domain.
Furthermore, the platform could include AI components or expert systems specific to the domain to support more realistic simulations.

\subsection{Immersive Experience}\label{immersive}
As discussed in Section~\ref{immersed}, we find that there is limited research on users' immersive experiences currently.
Popular science fiction TV series, \textit{Westworld}, set in a futuristic, highly immersive theme park populated by lifelike AI agents, which allows human guests to live out their fantasies in a Western-themed world without consequences.
As agent technology advances, the science fiction scenarios portrayed in the series are increasingly approaching reality.
Research on user immersive experience in ABMS is currently most relevant in the context of video games, such as role-playing games~(RPGs).
Värtinen\etal~\cite{c7c0852d5f324ba5907ee22bea26560c} generated role-playing game quests with LLMs to fulfill player demands toward more and richer game content.
By understanding how ABMS contributes to immersion, game developers can create environments that foster emotional investment, realistic social dynamics, and greater player satisfaction.
Additionally, insights gained may benefit other fields involving immersive environments, such as virtual reality.
Furthermore, as biotechnology and materials science advance to new levels, the concept of physical parks akin to \textit{Westworld} may become feasible.
Users would interact with physical agents through \textit{Natural Language} and \textit{Physical Movements}, creating highly immersive experiences.

Another potential application scenario is companion agents designed to provide emotional support. 
The rapid advancement of high technology has created a sense of disconnection and emotional distance, paradoxically leaving people feeling more alone despite constant virtual contact. 
Digital interactions often replace direct, face-to-face connections.
An inner emptiness or emotional void emerges, leading to a growing need for meaningful interaction and companionship.
These agents could offer companionship, simulate meaningful conversations, and respond empathetically to users' needs.
This application requires careful attention to emotional intelligence, personalization, and ethical considerations to ensure that the agents are both supportive and safe for users.
We believe that the user immersive experience in ABMS holds significant research value.

\section{Discussion}
In this section, we discuss some lessons learned during our work. 
We first introduce the trust issue and ethical problem arising from human-AI interactions.
We further discuss the future relationship between humans and AI.
It is hoped that this will stimulate further reflection among researchers.

\subsection{Trust Issue}
While LLMs offer many conveniences for interaction methods, they also introduce potential risks, such as the issue of ``hallucinations''~\cite{yao2024llmlieshallucinationsbugs}. 
This phenomenon occurs when the model generates inaccurate or misleading information with high confidence. It can undermine the reliability of ABMS outcomes, especially in critical applications.
Inspired by the algorithmic fidelity criteria proposed by Argyle\etal~\cite{Argyle_Busby_Fulda_Gubler_Rytting_Wingate_2023}, we have concluded three kinds of ``hallucinations'' in ABMS: 1)  generated outputs are distinguishable from parallel humans; 2) generated outputs are inconsistent with the predefined demographic information of agents; 3) generated outputs proceed unnaturally from the form, tone, and content of the context provided.
As a result, humans may experience trust issues with AI-generated outputs, which could pose risks for subsequent applications.
Therefore, exploring how human-AI interactions can mitigate the impact of hallucinations generated by LLMs can also be an important area of research.
For example, designing interactive mechanisms that allow users to verify, correct, or override misleading responses in real time could enhance the reliability of LLMs. 
Additionally, integrating feedback loops where users can flag inconsistencies or request clarifications may help manage and reduce the influence of hallucinations in critical ABMS applications.
On the other hand, designing appropriate mechanisms for LLMs to display their reasoning process transparently can enhance human trust.
Users can better grasp how conclusions are drawn and how outputs are generated. 
This transparency can help mitigate skepticism and uncertainty, allowing users to assess the model's logic and reliability more effectively.

\subsection{Ethical Problem}
Ethical problems arising from human-AI interactions in ABMS are a significant concern. 
Identifying ethical issues and exploring solutions is crucial in the field of HCI.
We provide two examples for reference as follows.
First, some ABMS rely on detailed data about individual demographics uploaded by users, especially in fields such as healthcare, urban planning, or the social sciences.
Using personal or sensitive data can risk breaching individuals' privacy if not handled securely or anonymized properly.
It is essential to use privacy-preserving techniques and comply with data protection laws to prevent unauthorized data access or misuse.
Comprehensive protection mechanisms need to be established to safeguard privacy and ensure the secure handling of sensitive data, ensure ethical use and transparency.
Second, simulated behaviors may inadvertently perpetuate biases and stereotypes embedded in LLMs' training data.
The training dataset may incorporate biases related to race, gender, ethnicity, and other characteristics~\cite{lucy-bamman-2021-gender}.
As a result, ethical considerations require researchers to take an active role in mitigating these biases.
Nonetheless, thoroughly identifying and mitigating all potential biases and stereotypes remains challenging, requiring continued research to further enhance and ensure the fairness of these models.

\subsection{Paradox of Coexist \textit{vs.} Compete}
\textit{``Carbon and Silicon, Coexist or Compete?''}, in the title, we raise the question of whether human~(carbon-based) and agent~(silicon-based) entities can coexist collaboratively or are destined to compete within shared environments in the future.
As generative AI systems demonstrate unprecedented reasoning, creativity, and autonomous decision-making capabilities, critical questions emerge: will humans and agents evolve as collaborative partners, or will their interactions devolve into zero-sum competition?
Modern AI exhibits dual potential as both ``augmenters'' and ``displacers'' of human capabilities.
It demonstrates how AI can amplify professional productivity while simultaneously threatening current occupations. 
Nevertheless, we think the human-AI relationship transcends binary competition or cooperation dichotomies, evolving instead as a ``recursive partnership'' where each entity redefines the other's capabilities.
In our paper, we examine diverse types of interactive modes between humans and agents, encompassing both egalitarian and hierarchical dynamics, as well as collaborative and directive forms of engagement.
The decisive factor is to implement adaptive governance frameworks that align AI's emergent properties with anthropogenic values.
Humans must establish clear boundaries, accountability frameworks, and trust mechanisms to ensure AI is used responsibly and beneficially.
The future relationship between humans and AI remains uncertain. 
Through our discussion of interactions in  ABMS, we aim to offer a perspective that may guide future researchers in exploring this evolving dynamic.
\section{Conclusion}
We conduct a systematic survey of 97 research studies on human-AI interactions in agent-based modeling and simulation in various domains from 1996 to 2024.
We first propose a novel taxonomy to categorize the interactions extracted from collected works.
We decompose each interaction into five dimensions according to the ``5W1H'' guideline.
Specially, we employ an analogy from the field of theater and draw upon some related professions to correspond to the roles of users.
Through our analysis, we answered the research question: \textit{How do humans and AI interact in the context of ABMS to fulfill user research requirements?}
Furthermore, we synthesize findings from existing literature to uncover interaction patterns, identify research gaps, and propose future research directions for human-AI interactions in agent-based modeling and simulation.

\bibliographystyle{ACM-Reference-Format}
\bibliography{sample-base}

\appendix

\section{Appendix}
\subsection{}\label{Ainitial}

\begin{longtable}{>{\arraybackslash}lp{2.7cm}p{0.8cm}llp{3.5cm}p{2cm}}
\caption[Short Caption]{This table introduces works concerning \textit{Initialize the Simulation}.}
\label{tab:initial} \\

\hline \textbf{Year}&\textbf{Work} & \textbf{Env}&\textbf{When} & \textbf{Who} & \textbf{What} & \textbf{How} \\  \hline 
\endfirsthead

\multicolumn{7}{c}%
{{\bfseries \tablename\ \thetable{} -- continued from previous page}} \\
\hline \textbf{Year}&\textbf{Work} & \textbf{Env}&\textbf{When} & \textbf{Who} & \textbf{What} & \textbf{How} \\  \hline  
\endhead

\hline \multicolumn{7}{r}{{Continued on next page}} \\ \hline
\endfoot

\hline \hline
\endlastfoot
\specialrule{0em}{1pt}{1pt}
2023&Generative Agents~\cite{10.1145/3586183.3606763}&S-P & Pre-S & Scriptwriter   &Agent~\includegraphics[width=0.025\textwidth]{icon/identity.pdf}&Language \\

2022&Social Simulacra~\cite{10.1145/3526113.3545616}&S-V & Pre-S & Scriptwriter   &Agent~\includegraphics[width=0.025\textwidth]{icon/identity.pdf}~\includegraphics[width=0.025\textwidth]{icon/goal.pdf}; Env~\includegraphics[width=0.025\textwidth]{icon/rule.pdf}&Language \\

2023&ChatEval~\cite{chan2023chatevalbetterllmbasedevaluators}&None & Pre-S & Scriptwriter   &Agent~\includegraphics[width=0.025\textwidth]{icon/identity.pdf}~\includegraphics[width=0.025\textwidth]{icon/learning.pdf} &Configuration \\

2023&MetaGPT~\cite{hong2024metagptmetaprogrammingmultiagent}&None & Pre-S & Scriptwriter   &Agent~\includegraphics[width=0.025\textwidth]{icon/goal.pdf}&Language \\

2023&Argyle\etal~\cite{Argyle_Busby_Fulda_Gubler_Rytting_Wingate_2023}&R-P & Pre-S & Prototype   &Agent~\includegraphics[width=0.025\textwidth]{icon/identity.pdf}&Data \\

2023&SayPlan~\cite{rana2023sayplangroundinglargelanguage}&S-P & Pre-S & Director   &Agent~\includegraphics[width=0.025\textwidth]{icon/goal.pdf}&Language \\

2023&AgentSims~\cite{lin2023agentsimsopensourcesandboxlarge}&S-P & Pre-S & Scriptwriter   &Agent~\includegraphics[width=0.025\textwidth]{icon/identity.pdf}~\includegraphics[width=0.025\textwidth]{icon/learning.pdf}; Env~\includegraphics[width=0.025\textwidth]{icon/object.pdf}&Interface; Configuration \\
2022&Huang\etal~\cite{huang2022innermonologueembodiedreasoning}&S-P; R-P & Pre-S & Director   &Agent~\includegraphics[width=0.025\textwidth]{icon/goal.pdf}&Language \\

2023&$S^3$~\cite{gao2023s3socialnetworksimulationlarge}&S-V & Pre-S & Prototype   &Agent~\includegraphics[width=0.025\textwidth]{icon/identity.pdf}&Data \\

2023&Ahn\etal~\cite{ahn2022icanisay}&R-P & Pre-S & Director   &Agent~\includegraphics[width=0.025\textwidth]{icon/goal.pdf}&Language \\

2022&WebShop~\cite{NEURIPS2022_82ad13ec}&S-V & Pre-S & Director   &Agent~\includegraphics[width=0.025\textwidth]{icon/goal.pdf}&Language \\

2023&Mind2Web~\cite{NEURIPS2023_5950bf29}&R-V & Pre-S & Director   &Agent~\includegraphics[width=0.025\textwidth]{icon/goal.pdf}&Language \\

2023&CAMEL~\cite{NEURIPS2023_a3621ee9}&None & Pre-S & Director   &Agent~\includegraphics[width=0.025\textwidth]{icon/goal.pdf}&Language \\

2023&Aher\etal~\cite{pmlr-v202-aher23a}&R-P & Pre-S & Prototype   &Agent~\includegraphics[width=0.025\textwidth]{icon/identity.pdf}&Data \\

2023&CGMI~\cite{jinxin2023cgmiconfigurablegeneralmultiagent}&S-P & Pre-S & Scriptwriter   &Env~\includegraphics[width=0.025\textwidth]{icon/description.pdf}&Language \\

2023&ChatLaw~\cite{cui2024chatlawmultiagentcollaborativelegal}&None & Pre-S & Director   &Agent~\includegraphics[width=0.025\textwidth]{icon/goal.pdf}&Language \\

2020&Alfred~\cite{shridhar2020alfredbenchmarkinterpretinggrounded}&R-P & Pre-S & Director   &Agent~\includegraphics[width=0.025\textwidth]{icon/goal.pdf}&Language \\
\specialrule{0em}{1pt}{1pt}
2023&Ren\etal~\cite{ren2023robotsaskhelpuncertainty}&R-P & Pre-S & Director   &Agent~\includegraphics[width=0.025\textwidth]{icon/goal.pdf}&Language \\

2023&ChatDev~\cite{qian2024chatdevcommunicativeagentssoftware}&None & Pre-S & Director   &Agent~\includegraphics[width=0.025\textwidth]{icon/goal.pdf}&Language \\

-&BactoWars~\cite{berryman2008review}&S-P & Pre-S & Scriptwriter  &Agent~\includegraphics[width=0.025\textwidth]{icon/identity.pdf}~\includegraphics[width=0.025\textwidth]{icon/interaction.pdf}; Env~\includegraphics[width=0.025\textwidth]{icon/object.pdf}&Configuration \\

-&EINSTein~\cite{berryman2008review}&S-P & Pre-S & Scriptwriter  &Agent~\includegraphics[width=0.025\textwidth]{icon/identity.pdf}~\includegraphics[width=0.025\textwidth]{icon/interaction.pdf}; Env~\includegraphics[width=0.025\textwidth]{icon/description.pdf}; Sim~\includegraphics[width=0.025\textwidth]{icon/technique.pdf}&Configuration \\

-&MANA~\cite{berryman2008review} &S-P & Pre-S & Scriptwriter  &Env~\includegraphics[width=0.025\textwidth]{icon/description.pdf}; Sim~\includegraphics[width=0.025\textwidth]{icon/technique.pdf}&Interface \\

2005&MASON~\cite{doi:10.1177/0037549705058073}&S-P & Pre-S & Scriptwriter   &Agent~\includegraphics[width=0.025\textwidth]{icon/identity.pdf}~\includegraphics[width=0.025\textwidth]{icon/learning.pdf}; Sim~\includegraphics[width=0.025\textwidth]{icon/technique.pdf}&Interface \\

1999&NetLogo~\cite{netlogo}&S-P & Pre-S & Scriptwriter  &Agent~\includegraphics[width=0.025\textwidth]{icon/identity.pdf}~\includegraphics[width=0.025\textwidth]{icon/interaction.pdf}&Interface \\

2006&North\etal~\cite{10.1145/1122012.1122013}&S-P & Pre-S & Scriptwriter  &Agent~\includegraphics[width=0.025\textwidth]{icon/identity.pdf}~\includegraphics[width=0.025\textwidth]{icon/learning.pdf}; Sim~\includegraphics[width=0.025\textwidth]{icon/technique.pdf} &Interface; Configuration \\

1996&Minar\etal~\cite{minar1996swarm}&S-P & Pre-S & Scriptwriter  &Agent~\includegraphics[width=0.025\textwidth]{icon/identity.pdf}~\includegraphics[width=0.025\textwidth]{icon/interaction.pdf}; Env~\includegraphics[width=0.025\textwidth]{icon/object.pdf}~\includegraphics[width=0.025\textwidth]{icon/rule.pdf} &Configuration \\

2023&ComplexWorld~\cite{basavatia2023complexworld}
&S-V & Pre-S & Scriptwriter  &Env~\includegraphics[width=0.025\textwidth]{icon/description.pdf}~\includegraphics[width=0.025\textwidth]{icon/object.pdf}~\includegraphics[width=0.025\textwidth]{icon/rule.pdf} &Language \\

2024&Cui\etal~\cite{Cui_2024_WACV}
&S-P & Pre-S & Director  &Agent~\includegraphics[width=0.025\textwidth]{icon/goal.pdf}&Language \\

2013&Gaube\etal~\cite{GAUBE201392}
&R-P & Pre-S & Prototype  &Agent~\includegraphics[width=0.025\textwidth]{icon/identity.pdf}&Data \\

2023&WarAgent~\cite{hua2024warpeacewaragentlarge}
&S-P & Pre-S & Scriptwriter  &Agent~\includegraphics[width=0.025\textwidth]{icon/identity.pdf}~\includegraphics[width=0.025\textwidth]{icon/interaction.pdf}; Env~\includegraphics[width=0.025\textwidth]{icon/description.pdf}~\includegraphics[width=0.025\textwidth]{icon/rule.pdf}&Configuration \\

2018&Kavak\etal~\cite{10.5555/3213032.3213044}
&S-P & Pre-S & Prototype  &Agent~\includegraphics[width=0.025\textwidth]{icon/identity.pdf}&Data \\

2023&Modelscope-agent~\cite{li2023modelscopeagentbuildingcustomizableagent}
&R-V & Pre-S & Scriptwriter  &Agent~\includegraphics[width=0.025\textwidth]{icon/identity.pdf}~\includegraphics[width=0.025\textwidth]{icon/learning.pdf}&Configuration \\

2024&AgentCoord~\cite{pan2024agentcoordvisuallyexploringcoordination}
&S-P & Pre-S & Director  &Agent~\includegraphics[width=0.025\textwidth]{icon/goal.pdf}&Interface \\

& & & & Scriptwriter  &Agent~\includegraphics[width=0.025\textwidth]{icon/identity.pdf}~\includegraphics[width=0.025\textwidth]{icon/interaction.pdf} &Interface \\

2023&Choicemates~\cite{park2023choicematessupportingunfamiliaronline}
&None & Pre-S & Director  &Agent~\includegraphics[width=0.025\textwidth]{icon/goal.pdf}; Env~\includegraphics[width=0.025\textwidth]{icon/description.pdf}&Language; Interface \\

2024&Rehearsal~\cite{10.1145/3613904.3642159}
&None & Pre-S & Director  &Env~\includegraphics[width=0.025\textwidth]{icon/description.pdf}&Interface \\

2023&Wang\etal~\cite{wang2024userbehaviorsimulationlarge}
&S-V & Pre-S & Scriptwriter  &Agent~\includegraphics[width=0.025\textwidth]{icon/identity.pdf}&Configuration \\

2023&Humanoid Agents~\cite{wang2023humanoidagentsplatformsimulating}
&S-P & Pre-S & Scriptwriter  &Agent~\includegraphics[width=0.025\textwidth]{icon/identity.pdf}&Language \\

2023&Zhu\etal~\cite{zhu2023ghostminecraftgenerallycapable}
&R-V & Pre-S & Director  &Agent~\includegraphics[width=0.025\textwidth]{icon/goal.pdf}&Language \\

-&PedSim~\cite{Ped} &S-P & Pre-S & Scriptwriter  &Agent~\includegraphics[width=0.025\textwidth]{icon/identity.pdf}~\includegraphics[width=0.025\textwidth]{icon/goal.pdf}; Env~\includegraphics[width=0.025\textwidth]{icon/object.pdf}~\includegraphics[width=0.025\textwidth]{icon/rule.pdf}&Configuration \\

-&AnyLogic~\cite{doi:https://doi.org/10.1002/9781118762745.ch12} &S-P & Pre-S & Scriptwriter  &Agent~\includegraphics[width=0.025\textwidth]{icon/identity.pdf}~\includegraphics[width=0.025\textwidth]{icon/goal.pdf}; Env~\includegraphics[width=0.025\textwidth]{icon/object.pdf}~\includegraphics[width=0.025\textwidth]{icon/rule.pdf}&Configuration \\

2023 &AutoGPT~\cite{Significant_Gravitas_AutoGPT} &None & Pre-S & Scriptwriter  &Agent~\includegraphics[width=0.025\textwidth]{icon/identity.pdf}&Configuration; Interface \\

2023 &BabyAGI~\cite{babyagi} &None & Pre-S & Scriptwriter  &Agent~\includegraphics[width=0.025\textwidth]{icon/identity.pdf}&Configuration \\

2023 &CommunityBots~\cite{10.1145/3579469} &None & Pre-S & Director  &Agent~\includegraphics[width=0.025\textwidth]{icon/goal.pdf}&Language; Interface \\
\specialrule{0em}{1pt}{1pt}
2024 &ComPeer~\cite{10.1145/3654777.3676430} &None & Pre-S & Director  &Agent~\includegraphics[width=0.025\textwidth]{icon/action.pdf}&Language; Interface \\

2024 &PrISM-Observer~\cite{arakawa2024prismobserverinterventionagenthelp} &R-P & Pre-S & Director  &Agent~\includegraphics[width=0.025\textwidth]{icon/action.pdf}&Physical \\

2024 &Jaber\etal~\cite{10.1145/3613904.3642183} &R-P & Pre-S & Director  &Agent~\includegraphics[width=0.025\textwidth]{icon/action.pdf}&Physical \\

2024 &Wan\etal~\cite{10.1145/3613905.3651026} &S-P & Pre-S & Director  &Agent~\includegraphics[width=0.025\textwidth]{icon/action.pdf}&Language; Physical \\

2024 &Chen\etal~\cite{10.1145/3613904.3642377} &None & Pre-S & Director  &Agent~\includegraphics[width=0.025\textwidth]{icon/action.pdf}&Interface \\

2024 &Zhang\etal~\cite{10.1145/3613904.3642545} &R-V & Pre-S & Director  &Agent~\includegraphics[width=0.025\textwidth]{icon/action.pdf}&Interface \\

2024 &Text2AC~\cite{10.1145/3613905.3651049} &R-V & Pre-S & Director  &Agent~\includegraphics[width=0.025\textwidth]{icon/identity.pdf}&Language; Interface \\
2023 &Ross\etal~\cite{10.1145/3581641.3584037} &None & Pre-S & Director  &Agent~\includegraphics[width=0.025\textwidth]{icon/goal.pdf}&Language; Interface \\
2024& ChatCam~\cite{10.1145/3699731}&R-P&Pre-S&Director&Agent~\includegraphics[width=0.025\textwidth]{icon/goal.pdf}&Language\\
2024& DrHouse~\cite{10.1145/3699765}&R-P&Pre-S&Director&Agent~\includegraphics[width=0.025\textwidth]{icon/goal.pdf}&Language\\
2024&ChatIoT~\cite{10.1145/3678585}&R-P&Pre-S&Director&Agent~\includegraphics[width=0.025\textwidth]{icon/goal.pdf}&Language\\
2024&CrowdBot~\cite{10.1145/3659601}&R-P&Pre-S&Director&Agent~\includegraphics[width=0.025\textwidth]{icon/goal.pdf}&Language\\
\end{longtable}

\subsection{}\label{Arefine}

\begin{longtable}{>{\arraybackslash}llp{0.8cm}lllp{1.5cm}}
\caption[Short Caption]{This table introduces works concerning \textit{Refine the Model}.}
\label{tab:refine} \\

\hline \textbf{Year}&\textbf{Work} & \textbf{Env}&\textbf{When} & \textbf{Who} & \textbf{What} & \textbf{How} \\  \hline 
\endfirsthead

\multicolumn{7}{c}%
{{\bfseries \tablename\ \thetable{} -- continued from previous page}} \\
\hline \textbf{Year}&\textbf{Work} & \textbf{Env}&\textbf{When} & \textbf{Who} & \textbf{What} & \textbf{How} \\  \hline  
\endhead

\hline \multicolumn{7}{r}{{Continued on next page}} \\ \hline
\endfoot

\hline \hline
\endlastfoot
\specialrule{0em}{1pt}{1pt}
2022 & Social Simulacra~\cite{10.1145/3526113.3545616} & S-V & D-S  & Scriptwriter    & Agent~\includegraphics[width=0.025\textwidth]{icon/goal.pdf}; Env~\includegraphics[width=0.025\textwidth]{icon/rule.pdf} & Language      \\
2012 & Prom Week~\cite{10.1145/2282338.2282384} & S-P & D-S  & Director    & Agent~\includegraphics[width=0.025\textwidth]{icon/action.pdf} & Interface     \\
2011 & Prom Week~\cite{10.1145/2159365.2159425} & S-P & D-S  & Director    & Agent~\includegraphics[width=0.025\textwidth]{icon/action.pdf} & Interface     \\
2023 & AGENTVERSE~\cite{chen2023agentversefacilitatingmultiagentcollaboration} & S-P & D-S  & Director    & Sim~\includegraphics[width=0.025\textwidth]{icon/progress.pdf} & Interface     \\
- & - & - & D-S  & Director    & Agent~\includegraphics[width=0.025\textwidth]{icon/goal.pdf} & Language      \\
2023 & ChatEval~\cite{chan2023chatevalbetterllmbasedevaluators} & None & D-S  & Director    & Agent~\includegraphics[width=0.025\textwidth]{icon/goal.pdf}; Sim~\includegraphics[width=0.025\textwidth]{icon/progress.pdf} &   Language; Interface     \\

2024 & Zhang\etal~\cite{zhang2024buildingcooperativeembodiedagents} & S-P & D-S  & Actor    & Agent~\includegraphics[width=0.025\textwidth]{icon/action.pdf} &   Language; Interface     \\
2023 & Memory sandbox~\cite{10.1145/3586182.3615796} & None & D-S  & Scriptwriter    & Agent~\includegraphics[width=0.025\textwidth]{icon/learning.pdf} & Interface     \\
2022 & Inner monologue~\cite{huang2022innermonologueembodiedreasoning} & S-P; R-P & D-S  & Director    & Agent~\includegraphics[width=0.025\textwidth]{icon/goal.pdf} & Language      \\
2022 & Krishna\etal~\cite{doi:10.1073/pnas.2115730119} & R-V & Pre-S   & Director    & Agent~\includegraphics[width=0.025\textwidth]{icon/learning.pdf} & Language      \\
2023 & RoCo~\cite{mandi2023rocodialecticmultirobotcollaboration} & R-P & D-S  & Actor    & Agent~\includegraphics[width=0.025\textwidth]{icon/action.pdf} & Physical; Language      \\
2023 & ChatLaw~\cite{cui2024chatlawmultiagentcollaborativelegal} & None & D-S  & Director    & Agent~\includegraphics[width=0.025\textwidth]{icon/learning.pdf} & Language      \\
\specialrule{0em}{1pt}{1pt}
2023 & Mehta\etal~\cite{mehta2024improvinggroundedlanguageunderstanding} & S-P & D-S  & Director    & Agent~\includegraphics[width=0.025\textwidth]{icon/action.pdf} & Language      \\
\specialrule{0em}{1pt}{0pt}
2020 & Alfred~\cite{shridhar2020alfredbenchmarkinterpretinggrounded} & R-P & D-S  & Director    & Agent~\includegraphics[width=0.025\textwidth]{icon/action.pdf} & Language      \\
2023 & Mohanty\etal~\cite{mohanty2023transforminghumancenteredaicollaboration} & S-P & D-S  & Director    & Agent~\includegraphics[width=0.025\textwidth]{icon/action.pdf} & Language      \\
2022 & Teach~\cite{Padmakumar_Thomason_Shrivastava_Lange_Narayan-Chen_Gella_Piramuthu_Tur_Hakkani-Tur_2022} & R-P & D-S  & Director    & Agent~\includegraphics[width=0.025\textwidth]{icon/action.pdf} & Language      \\
2023& Ren\etal~\cite{ren2023robotsaskhelpuncertainty} & R-P & D-S  & Director    & Agent~\includegraphics[width=0.025\textwidth]{icon/action.pdf} & Language      \\
2005 & MASON~\cite{doi:10.1177/0037549705058073} & S-P & D-S  & Director    & Sim~\includegraphics[width=0.025\textwidth]{icon/progress.pdf} & None \\
2006 & North\etal~\cite{10.1145/1122012.1122013} & S-P & D-S  & Director    & Sim~\includegraphics[width=0.025\textwidth]{icon/progress.pdf} & Interface     \\
2023 & Drive like a human~\cite{unknown} & R-P & D-S  & Director    & Agent~\includegraphics[width=0.025\textwidth]{icon/learning.pdf} & Language      \\
2006 & Guyot~\etal~\cite{guyot2006} & R-V & D-S  & Actor    & Agent~\includegraphics[width=0.025\textwidth]{icon/action.pdf} & Interface     \\
2023 & Surrealdriver~\cite{jin2024surrealdriverdesigningllmpoweredgenerative} & S-P & Pre-S   & Director    & Agent~\includegraphics[width=0.025\textwidth]{icon/learning.pdf} & Data   \\
2023 & The SocialAI School~\cite{kovač2023socialaischoolinsightsdevelopmental} & S-P & Pre-S   & Director    & Agent~\includegraphics[width=0.025\textwidth]{icon/learning.pdf} & Interface     \\
2023 & Choicemates~\cite{park2023choicematessupportingunfamiliaronline} & None & D-S  & Director    & Agent~\includegraphics[width=0.025\textwidth]{icon/action.pdf}~\includegraphics[width=0.025\textwidth]{icon/interaction.pdf} &   Language; Interface     \\
2023 & Ghost in the minecraft~\cite{zhu2023ghostminecraftgenerallycapable} & R-V & D-S  & Director    & Agent~\includegraphics[width=0.025\textwidth]{icon/action.pdf} & Language      \\
2024 & Lu\etal~\cite{10.1145/3672539.3686351} & None & D-S  & Director    & Agent~\includegraphics[width=0.025\textwidth]{icon/action.pdf} & Interface     \\
2024 & Teach AI How to Code~\cite{10.1145/3613904.3642349} & None & D-S  & Director    & Agent~\includegraphics[width=0.025\textwidth]{icon/learning.pdf} &   Language; Interface     \\
2024 & Zhou\etal~\cite{10.1145/3613904.3642812} & None & D-S  & Director    & Agent~\includegraphics[width=0.025\textwidth]{icon/goal.pdf} & Language      \\
2022 & Wordcraft~\cite{10.1145/3490099.3511105} & None & D-S  & Director    & Agent~\includegraphics[width=0.025\textwidth]{icon/goal.pdf} & Interface     \\
2023 & Ross~\etal~\cite{10.1145/3581641.3584037} & None & D-S  & Director    & Sim~\includegraphics[width=0.025\textwidth]{icon/progress.pdf} & Interface    
\end{longtable}
\subsection{}\label{Aevaluate}

\begin{longtable}{>{\arraybackslash}lp{3.1cm}llllp{1.8cm}}
\caption[Short Caption]{This table introduces works concerning \textit{Evaluate the Performance}.}
\label{tab:evaluate} \\

\hline \textbf{Year}&\textbf{Work} & \textbf{Env}&\textbf{When} & \textbf{Who} & \textbf{What} & \textbf{How} \\  \hline 
\endfirsthead

\multicolumn{7}{c}%
{{\bfseries \tablename\ \thetable{} -- continued from previous page}} \\
\hline \textbf{Year}&\textbf{Work} & \textbf{Env}&\textbf{When} & \textbf{Who} & \textbf{What} & \textbf{How} \\  \hline  
\endhead

\hline \multicolumn{7}{r}{{Continued on next page}} \\ \hline
\endfoot

\hline \hline
\endlastfoot
\specialrule{0em}{1pt}{1pt}
2023 & Generative Agents~\cite{10.1145/3586183.3606763} & S-P & Post-S   & Actor   & Agent~\includegraphics[width=0.025\textwidth]{icon/action.pdf}~\includegraphics[width=0.025\textwidth]{icon/identity.pdf}~\includegraphics[width=0.025\textwidth]{icon/learning.pdf} & Language\\
- & - & - & Post-S   & Observer   & Agent~\includegraphics[width=0.025\textwidth]{icon/action.pdf} & Data   \\
2022 & Social Simulacra~\cite{10.1145/3526113.3545616} & S-V & Pre-S   & Scriptwriter   & Agent~\includegraphics[width=0.025\textwidth]{icon/goal.pdf}; Env~\includegraphics[width=0.025\textwidth]{icon/rule.pdf} & Language\\
- & - & - & Post-S   & Observer   & Agent~\includegraphics[width=0.025\textwidth]{icon/action.pdf} & Data   \\
2024 & Zhang\etal~\cite{zhang2024buildingcooperativeembodiedagents} & S-P & D-S  & Actor   & Agent~\includegraphics[width=0.025\textwidth]{icon/action.pdf} & Language; Interface   \\
2023 & MetaGPT~\cite{hong2024metagptmetaprogrammingmultiagent} & None & Post-S   & Observer   & Agent~\includegraphics[width=0.025\textwidth]{icon/action.pdf} & Interface   \\
2023 & Argyle\etal~\cite{Argyle_Busby_Fulda_Gubler_Rytting_Wingate_2023} & R-P & Post-S   & Observer   & Agent~\includegraphics[width=0.025\textwidth]{icon/action.pdf} & Data   \\
2023 & AgentSims~\cite{lin2023agentsimsopensourcesandboxlarge} & S-P & D-S  & Observer   & Agent~\includegraphics[width=0.025\textwidth]{icon/action.pdf} & Interface   \\
2023 & Saha\etal~\cite{saha2023languagemodelsteachweaker} & None & D-S  & Actor   & Agent~\includegraphics[width=0.025\textwidth]{icon/action.pdf} & Language\\
\specialrule{0em}{1pt}{1pt}
2023 & CAMEL~\cite{NEURIPS2023_a3621ee9} & None & Post-S   & Observer   & Agent~\includegraphics[width=0.025\textwidth]{icon/action.pdf} & Data   \\
-& BactoWars~\cite{berryman2008review} & S-P & Post-S   & Observer   & Agent~\includegraphics[width=0.025\textwidth]{icon/action.pdf} & Interface   \\
2022 & FAIR\etal~\cite{doi:10.1126/science.ade9097} & R-V & D-S  & Actor   & Agent~\includegraphics[width=0.025\textwidth]{icon/action.pdf} & Interface   \\
2020 & Feng\etal~\cite{10.1145/3394486.3412862} & R-P & Pre-S   & Prototype   & Agent~\includegraphics[width=0.025\textwidth]{icon/identity.pdf} & Data   \\
2023 & H\"{a}m\"{a}l\"{a}inen\etal~\cite{10.1145/3544548.3580688} & R-P & Post-S   & Observer   & Agent~\includegraphics[width=0.025\textwidth]{icon/action.pdf} & Data   \\
2023 & War and Peace~\cite{hua2024warpeacewaragentlarge} & S-P & Post-S   & Observer   & Agent~\includegraphics[width=0.025\textwidth]{icon/action.pdf} & Data   \\
2023 & Surrealdriver~\cite{jin2024surrealdriverdesigningllmpoweredgenerative} & S-P & Post-S   & Observer   & Agent~\includegraphics[width=0.025\textwidth]{icon/action.pdf} & Data   \\
2023 & Modelscope-agent~\cite{li2023modelscopeagentbuildingcustomizableagent} & R-V & Pre-S   & Observer   & Agent~\includegraphics[width=0.025\textwidth]{icon/action.pdf} & Interface   \\
2023 & Liu\etal~\cite{liu2023trainingsociallyalignedlanguage} & S-P & Post-S   & Observer   & Agent~\includegraphics[width=0.025\textwidth]{icon/action.pdf} & Data   \\
2023 & Alympics~\cite{mao2024alympicsllmagentsmeet} & S-V & Post-S   & Observer   & Agent~\includegraphics[width=0.025\textwidth]{icon/action.pdf} & Data   \\
2024 & AgentCoord~\cite{pan2024agentcoordvisuallyexploringcoordination} & S-P & D-S  & Director   & Agent~\includegraphics[width=0.025\textwidth]{icon/action.pdf} & Interface   \\
2023 & Choicemates~\cite{park2023choicematessupportingunfamiliaronline} & None & D-S  & Observer   & Agent~\includegraphics[width=0.025\textwidth]{icon/action.pdf} & Interface   \\
- & - & - & Pre-S   & Director   & Agent~\includegraphics[width=0.025\textwidth]{icon/goal.pdf} & Language; Interface   \\
- & - & - & Post-S   & Observer   & Agent~\includegraphics[width=0.025\textwidth]{icon/action.pdf} & Data   \\
2024 & Schwitzgebel\etal~\cite{https://doi.org/10.1111/mila.12466} & None & Post-S   & Observer   & Agent~\includegraphics[width=0.025\textwidth]{icon/action.pdf} & Data   \\
2024 & Rehearsal~\cite{10.1145/3613904.3642159} & None & D-S  & Director   & Agent~\includegraphics[width=0.025\textwidth]{icon/action.pdf} & Interface   \\
2023 & Wang\etal~\cite{wang2024userbehaviorsimulationlarge} & S-V & Post-S   & Observer   & Agent~\includegraphics[width=0.025\textwidth]{icon/action.pdf} & Data   \\
2023 & Humanoid Agents~\cite{wang2023humanoidagentsplatformsimulating} & S-P & D-S  & Observer   & Agent~\includegraphics[width=0.025\textwidth]{icon/action.pdf} & Interface   \\
- & - & - & Post-S   & Observer   & Agent~\includegraphics[width=0.025\textwidth]{icon/action.pdf} & Data   \\
2023 & Zhang\etal~\cite{10.1145/3626772.3657844} & S-V & Pre-S   & Prototype   & Agent~\includegraphics[width=0.025\textwidth]{icon/identity.pdf} & Data   \\
2024 & SOTOPIA~\cite{zhou2024sotopiainteractiveevaluationsocial} & None & D-S  & Actor   & Agent~\includegraphics[width=0.025\textwidth]{icon/action.pdf} & Interface   \\
-& PedSim~\cite{Ped} & S-P & D-S  & Observer   & Agent~\includegraphics[width=0.025\textwidth]{icon/action.pdf} & Interface   \\
-& AnyLogic~\cite{doi:https://doi.org/10.1002/9781118762745.ch12} & S-P & D-S  & Observer   & Agent~\includegraphics[width=0.025\textwidth]{icon/action.pdf} & Interface   \\
-& AutoGPT~\cite{Significant_Gravitas_AutoGPT} & None & D-S  & Observer   & Agent~\includegraphics[width=0.025\textwidth]{icon/action.pdf} & Interface   \\
-& BabyAGI~\cite{babyagi} & None & D-S  & Observer   & Agent~\includegraphics[width=0.025\textwidth]{icon/action.pdf} & Data   \\
2021 & Siu\etal~\cite{NEURIPS2021_86e8f7ab} & R-V & D-S  & Actor   & Agent~\includegraphics[width=0.025\textwidth]{icon/action.pdf} & Interface   \\
2023 & Eloy\etal~\cite{10.1145/3579598} & S-P & D-S  & Actor   & Agent~\includegraphics[width=0.025\textwidth]{icon/action.pdf} & Language; Interface   \\
2023 & Zubatiy\etal~\cite{10.1145/3610170} & None & D-S  & Director   & Agent~\includegraphics[width=0.025\textwidth]{icon/action.pdf} & Language; Interface   \\
2024 & Jaber\etal~\cite{10.1145/3613904.3642183} & R-P & D-S  & Director   & Agent~\includegraphics[width=0.025\textwidth]{icon/action.pdf} & Physical \\
2024 & Dai\etal~\cite{10.1145/3613905.3637145} & S-P & D-S  & Actor   & Agent~\includegraphics[width=0.025\textwidth]{icon/action.pdf} & Physical; Language\\
2024 & Wan\etal~\cite{10.1145/3613905.3651026} & S-P & D-S  & Director   & Agent~\includegraphics[width=0.025\textwidth]{icon/action.pdf} & Language; Interface   \\
- & - & - & Post-S   & Observer   & Agent~\includegraphics[width=0.025\textwidth]{icon/action.pdf} & Data   \\
2024 & PeerGPT~\cite{10.1145/3613905.3651008} & R-P & D-S  & Actor   & Agent~\includegraphics[width=0.025\textwidth]{icon/action.pdf} & Language\\
2024 & ClassMeta~\cite{10.1145/3613904.3642947} & S-P & D-S  & Actor   & Agent~\includegraphics[width=0.025\textwidth]{icon/action.pdf} & Physical; Language\\
\specialrule{0em}{1pt}{1pt}
2024 & Attig\etal~\cite{10.1145/3613905.3650853} & R-V & D-S  & Actor   & Agent~\includegraphics[width=0.025\textwidth]{icon/action.pdf} & Interface   \\
2024 & Hwang\etal~\cite{10.1145/3613904.3642202} & R-P & D-S  & Director   & Agent~\includegraphics[width=0.025\textwidth]{icon/action.pdf} & Language\\
2024&DrHouse~\cite{10.1145/3699765}&R-P&Post-S&Observer&Agent~\includegraphics[width=0.025\textwidth]{icon/action.pdf}&Data\\
2024&Cuadra\etal~\cite{10.1145/3659624}&R-P&D-S&Director&Agent~\includegraphics[width=0.025\textwidth]{icon/goal.pdf}&Language; Interface\\
2024&Sasha~\cite{10.1145/3643505}&R-P&D-S&Director&Agent~\includegraphics[width=0.025\textwidth]{icon/goal.pdf}&Language\\
\end{longtable}

\end{document}